\documentclass[aps,prb,twocolumn,superscriptaddress,longbibliography]{revtex4-2}
\usepackage{amsmath,amssymb}
\usepackage[pdftex]{hyperref,graphicx}
\hypersetup{colorlinks = true, urlcolor = blue, linkcolor = blue, citecolor = blue}
\usepackage{physics}
\usepackage{xcolor}
\usepackage{bm}

\newcommand{\argmax}{\operatornamewithlimits{argmax}}

\begin{document}
\title{Majorana zero modes in semiconductor-superconductor hybrid structures: Defining topology in short and disordered nanowires through Majorana splitting}
\author{Haining Pan}
\affiliation{Department of Physics and Astronomy, Center for Materials Theory, Rutgers University, Piscataway, New Jersey 08854 USA}
\author{Sankar Das Sarma}
\affiliation{Condensed Matter Theory Center and Joint Quantum Institute, Department of Physics, University of Maryland, College Park, Maryland 20742, USA}

\begin{abstract}
Majorana zero modes (MZMs) are bound midgap topological excitations at the ends of a 1D topological superconductor, which must come in pairs. If the two MZMs in the pair are sufficiently well-separated by a distance much larger than their individual localization lengths, then the MZMs behave as non-Abelian anyons which can be braided to carry out fault-tolerant topological quantum computation.  
In this `topological' regime of well-separated MZMs, their overlap is exponentially small, leading to exponentially small Majorana splitting, thus enabling the MZMs to be topologically protected by the superconducting gap.  
In real experimental samples, however, the existence of disorder and the finite length of the 1D wire considerably complicate the situation, leading to ambiguities in defining `topology' since the Majorana splitting between the two end modes may not necessarily be small in disordered wires of short length. 
We theoretically study this situation by calculating the splitting in experimentally relevant short disordered wires, and explicitly investigating the applicability of the `exponential protection' constraint as a function of disorder, wire length, and other system parameters in realistic models of nanowires currently being used experimentally.  
We find that the exponential regime is highly constrained, and is suppressed for disorder somewhat less than the topological superconducting gap. 
We provide detailed results and discuss the implications of our theory for the currently active experimental search for MZMs in superconductor-semiconductor hybrid platforms. 
A general consequence of our work is that `topology' in finite disordered wires may not be uniquely defined, necessitating a careful analysis which depends on the context.
\end{abstract}

\maketitle

\section{BACKGROUND and INTRODUCTION}
In 2000, Kitaev pointed out~\cite{kitaev2001unpaired} that an infinitely long clean spinless $p$-wave superconducting 1D (lattice) system can be in two distinct phases: Trivial and topological, with a gapless quantum phase transition point separating the two. 
The topological phase has isolated midgap zero-energy Majorana bound states (MBS) localized at each end of the wire. It is `topological' since such isolated MBS are non-Abelian anyons manifesting intrinsic quantum degeneracy and nontrivial braiding statistics, which are protected by the bulk superconducting gap, as already pointed out by Read in the context of non-Abelian fractional quantum Hall states also in 2000.~\cite{read2000paired}  The trivial phase, by contrast, has no isolated MBS as the Majorana modes pair up to form regular fermionic states. The topological and trivial phases are separated by a gapless topological quantum phase transition (TQPT) in the parameter space of the Hamiltonian controlling the superconductivity. Kitaev also identified a `Pfaffian' topological indicator, which vanishes at the TQPT (where the superconducting gap must also vanish with the gap below and above the TQPT respectively being trivial and topological) which was also discussed by Read, taking on the values of $\pm 1$ in the trivial/topological phase. These early works, along with the 2000 work of Sengupta \textit{et al}.~\cite{sengupta2001midgap} demonstrating that such a topological spinless $p$-wave superconductor with isolated midgap MBS would manifest perfect Andreev reflection with a conductance quantization of $\frac{2e^2}{h}$, were done for pristine long wires, where topology is uniquely defined. (We note that it is well-known that such midgap zero-energy bound states cannot occur in regular $s$-wave superconductors by virtue of the fermion doubling theorem --- in $s$-wave superconductors, in-gap quasiparticles are the usual particle-hole symmetric electrons and holes with energies above or below the midgap.)

What happens to topology in finite and/or disordered systems? The important question of the operational definition of how these topological ideas developed for pristine and infinite systems apply to real experimental systems in the laboratory was not discussed at all in these early pioneering publications.~\cite{kitaev2001unpaired,read2000paired,sengupta2001midgap} A superconductor is uniquely defined by one energy scale, the gap ($\Delta$), and one length scale, the coherence length ($\xi$), and the obvious implication is that if the disorder strength ($\sigma$) and the wire length ($L$) are much smaller and much larger than $\Delta$ and $\xi$ respectively, the system is topological. For very short wires ($L\ll\xi$) and very strong disorder ($\Delta\ll \sigma$), topology cannot be defined, and there is no MBS or topological superconductivity. Quite coincidentally, again in 2000, another important (and independent) work~\cite{motrunich2001griffiths} showed that such topological $p$-wave superconductivity in an infinite 1D system is destroyed by strong disorder through a quantum phase transition taking the system from a topological superconductor at weak disorder to a trivial localized insulator at strong disorder. Thus, very strong disorder ($\Delta\ll\sigma$) necessarily destroys topology, although topology survives weak disorder. It is also clear that very short wires ($L\ll\xi$) cannot host isolated MBS since the two end MBS would overlap, creating ordinary fermionic Andreev bound states (ABS) shifted from zero energy, and thus, becoming nontopological. This shift in energy from zero energy corresponds to the `Majorana splitting' energy as the overlap between the two MBS at two ends split their zero-energy character, but for low enough overlap, i.e. when the two MBS are separated by $L\gg \xi$, the splitting is exponentially small $\sim \exp(-L/\xi)$, implying the topological protection of MBS. But this exponential topological protection applies only when $L\gg \xi$, and for short wires, topology becomes ill-defined. The same is true for strong disorder, where the length $L$ corresponds to the electron localization length, which is suppressed below the wire length, again leading to the effective short-wire situation since the localization length becomes comparable to or shorter than the coherence length $\xi$.

A key motivation of Ref.~\cite{kitaev2001unpaired} was the possibility of using the isolated end MBS for topological quantum computation~\cite{nayak2008nonabelian}, provided a 1D topological superconductor can be found. This can be done by suitably manipulating the MBS in their degenerate quantum subspace, defined by both MBS occupied or empty, as indicated by their parity, provided they are located sufficiently far from each other. In this scenario, the topological protection is `exponential', i.e., the overlap between the two MBS at the ends leads to an exponentially small energy splitting going as $\exp(-L/\xi)$  in the $L\gg\xi$ regime, thus preserving their zero energy status which is crucial for topology.   
Since in many situations, the coherence length (which in the topological superconducting state implies the size of the MBS localization at the end, thus controlling their end-to-end overlap suppressing the Majorana identity—for strong overlap the states are simply ABS), itself goes as $\frac{1}{\Delta}$. The superconducting gap $\Delta$ controls the topological protection—the larger the gap, the shorter the coherence length, and hence the stronger the protection through the $\exp(-L/\xi)$ factor in the Majorana splitting. However, such exponential regimes may not exist in short wires or even in longer wires with strong disorder and/or small energy gap. No direct experimental measurement of $\xi$ is feasible, so in realistic situations, we do not know if the $L\gg \xi$  topological condition is satisfied. Note that disorder may make $L$ much smaller than the wire length because the electron localization length is the relevant quantity to be compared with the SC coherence length, and for strong disorder the electron localization length may very well be shorter than the wire length in realistic devices.

The subject of MBS-carrying 1D topological superconductors remained academic and theoretical since nobody knew how to create such a system practically, and the early focus on non-Abelian anyons was entirely on the even-denominator fractional quantum Hall effect, where such quasiparticles should exist, but have not yet been observed because of various experimental difficulties.~\cite{dean2008intrinsic,nayak2008nonabelian}
In fact, the subject of MBS in 1D topological superconductors was essentially ignored for the first ten years up to 2010.  
This all changed in 2009-2010 with the appearance of several theoretical papers proposing a practical route to artificially engineer MBS-carrying semiconductor `Majorana nanowires' which can be tuned into the topological phase by varying an applied magnetic field and/or the chemical potential (through a suitable gate voltage).~\cite{lutchyn2010majorana,sau2010generic,sau2010nonabelian,oreg2010helical}
In this nanowire context, the MBS is typically called `Majorana zero modes' (MZM), and we adopt this terminology from now on in place of MBS. 
We restrict our attention mostly to the InAs/Al semiconductor-superconductor  (SM-SC) platform where InAs is the MZM-carrying nanowire in the topological phase (with Al providing the proximity-induced superconductivity in InAs, which is driven into the topological phase by the applied magnetic field). All our explicit results in this paper (except for a few, as explicitly noted at appropriate places) are given for the InAs/Al Majorana nanowire platform using physical units (i.e., meV, nm, T for energy, length, magnetic field, respectively) applicable to the experimental situations. It is easy to convert them into dimensionless units using the energy gap, the wire length (or the coherence length), etc., as the appropriate units.

In the context of the extensive experimental studies of Majorana nanowires~\cite{das2012zerobias,deng2012anomalous,mourik2012signatures,churchill2013superconductornanowire,finck2013anomalous,nichele2017scaling,zhang2017quantized,zhang2021large,song2021large,microsoftquantum2023inasal,aghaee2025interferometric} following the theoretical predictions, it became clear that the nanowires used in all the early experiments up to 2020 were both very short $(L< \xi)$ and very disordered ($\Delta<\sigma$), and the experimentally observed zero-bias conductance peaks (ZBCP) are simply low-energy fermionic Andreev bound states (ABS) accidentally localized near one wire end (but not both).~\cite{pan2020physical,dassarma2021disorderinduced,ahn2021estimating,dassarma2023search,kouwenhoven2025perspective,zhang2017quantized} 
It was shown explicitly using the random matrix theory that strongly disordered nanowires have a small but experimentally relevant generic probability of producing almost-zero-energy non-topological ABS near the wire ends, which could mimic MZMs,  thus leading to an erroneous identification of ABS as MBS.~\cite{pan2020generic} 
The experimentally studied systems were never in the topological phase, and the accidental ABS appeared through fine-tuning. A question then arises about the minimal necessary conditions for ascertaining the experimental MZM signatures, since the observation of ZBCP by itself is clearly insufficient. We have earlier studied this question of minimal experimental topological signatures in depth~\cite{pan2021threeterminal}, suggesting the necessary tunneling characteristics to be: (1) The observation of simultaneous ZBCP in local tunneling at both ends; (2) the observation of bulk gap closing/reopening signatures in nonlocal end-to-end tunneling precisely in the parameter regime where the ZBCP from the ends show up. Theoretical simulations can then further add sufficiency conditions by calculating the appropriate topological invariant, such as the Pfaffian or, more appropriately (for tunneling transport systems such as nanowires), the scattering matrix invariant (which is essentially equivalent to the Pfaffian invariant).~\cite{akhmerov2011quantized,fulga2011scattering,fulga2012scattering} The strictly theoretical binary values of the {topological invariant (TI)} are $\pm 1$ corresponding to trivial/topological regimes in the long wire ($L\gg \xi$, $\Delta\gg\sigma$) pristine limit, which is the limit where topology can be uniquely defined. At the TQPT, the TI, by definition, vanishes. Other idealized thermodynamic topological invariants share similar binary properties since idealized topology is binary: the system is either topological or not—this is essentially the cartoon example of a topological donut either having a hole/handle or not, and the only way to go from one to the other (i.e., from having a hole/handle to not having one) is to have a topological phase transition. For ideal systems, this definition is unique and precise.

Experimental systems are finite disordered nanowires, which may not be in the pristine long wire limit at all, and a serious problem is that there is no \textit{a priori} way of knowing if the conditions $L\gg\xi$, $\Delta\gg\sigma$ are satisfied in a given sample through any single direct measurement.
There is no \textit{a priori} way of determining whether the topology in a real system can even be defined, since the overlap between the end MZM is unknown and undetermined experimentally. In Ref.~\cite{dassarma2016how}, TI is generalized from a binary $\mathbb{Z}_2$ index (specifically $\pm 1$, with the TQPT being the zero value of the TI) to a continuous variable to accommodate realistic finite wires. In fact, recent theoretical work~\cite{dassarma2023spectral,dassarma2023density,pan2024disordered,day2025identifying} shows that in realistic wires currently being used even in the best experimental systems~\cite{microsoftquantum2023inasal}, the TI calculations are at best inconclusive since there could be many low-energy ABS along the wire in addition to any end-localized modes, and the TI may produce misleading conclusions simply because the topological invariants are designed only to detect the presence of TQPT in the long pristine limit, and they may not work in realistic wires because of their finite length and finite disorder. Since experimental tunneling spectroscopy has not yet seen the predicted robust conductance quantization, it remains unclear if the nanowires are topological or not, although a recent experiment has seen some signatures for the putative necessary conditions, i.e., ZBCP in local tunneling from both ends and gap closing/reopening in nonlocal tunneling from both ends. The extracted topological gap is, however, rather small ($\sim$ 10--30 $\mu$eV), and the topological regime in the experimental parameter space of gate voltage--magnetic field is rather patchy and small.~\cite{microsoftquantum2023inasal}

The unsatisfactory situation described above has created uncertainties about the existence of topology and non-Abelian MZM in experimental nanowires, although it is generally accepted that the theoretical SM-SC hybrid nanowires should host end MZM if the experimental wires are long enough and pristine enough. Given this context, in the current work, we take a different approach to understanding and analyzing topology in realistic SM-SC nanowires by going back to the fundamentals and defining the topology as achieving an exponential suppression in the MZM overlap as asserted in the original work of Kitaev. In fact, topological immunity is equivalent to the exponential suppression in the MZM energy splitting since expressing a fermion operator in terms of Majorana operators is an identity with no physical significance---the fundamental physics is in the corresponding MZMs being physically separated satisfying $L\gg\xi$ so that they are isolated. This isolation, manifested in their energy splitting going as $\sim \exp(-L/\xi)$, is the essence of topological immunity and the non-Abelian nature of MZM. In the absence of this exponential suppression, the MZMs combine, forming fermionic ABS, which certainly have some Majorana character because in some sense all subgap states in superconductors represent Majorana fermions because of the exact electron-hole symmetry of superconductors, but Majorana fermions by themselves are fermions and not topological non-Abelian anyons. What is needed for the existence of true non-Abelian MZM is the exponential suppression of the Majorana splitting, and this is the central theme of the current work. What is needed to achieve exponential suppression in Majorana nanowires in the presence of disorder and finite wire length? This is what we investigate in depth in the current work--- trying to tackle the difficult challenge of finding an operational working definition for topology in realistic systems with disorder and finite wire length.

Given that the Majorana splitting in the topological regime goes as $\sim \exp(-L/\xi)$ for $L\gg\xi$, it seems that all we need is to know the effective $L$ and effective $\xi$, and then the $L\gg\xi$ condition will ensure exponential suppression. There are, however, serious conceptual and practical problems in applying such a simple scenario to the realistic situations in the presence of disorder and finite wire length, even in the theoretical examples. For example, the coherence length $\xi$ is generally unknown, and $L$ may not, because of Anderson localization, necessarily be the mentioned wire length in the presence of disorder, as already discussed above.
Also, the \textit{in-situ} disorder strength in the nanowire is generally unknown since the system is a hybrid SM-SC structure, and measurement of the mobility in the starting semiconductor material may not provide an accurate estimate of the mean free path in the Majorana nanowire.

There is an additional conceptual problem associated with `Majorana oscillations' discovered in Refs.~\onlinecite{cheng2009splitting,cheng2010tunneling} and extensively discussed in the context of realistic Majorana nanowire physics in Ref.~\onlinecite{dassarma2012splitting}, along with other theory works~\cite{hegde2016majorana,penaranda2018quantifying,avila2020majorana,cayao2017majorana,zyuzin2013correlations}. The MZM splitting, even for long wires, does not simply decrease exponentially with increasing length, but also manifests an oscillatory behavior with the envelope of the oscillations decreasing exponentially. These Majorana (splitting) oscillations are related to well-known Friedel oscillations, and the splitting follows, ignoring some unimportant constant phase factor, a behavior $\sim \exp(-L/\xi)\cos (k_FL)$, where $k_F$ is an effective Fermi wavenumber associated with the MZM. Thus, increasing (decreasing) $L$ $(\xi)$ suppresses the oscillatory MZM splitting. By contrast, decreasing (increasing) $L$ ($\xi$) exponentially enhances the MZM splitting oscillations. In the trivial phase below TQPT, there are no MZM splitting oscillations as there are no MZMs in the trivial phase. Therefore, Kitaev's simple prescription for topological immunity, being synonymous with an exponential $\exp(-L/\xi)$ dependence of the MZM splitting, cannot be straightforwardly applied to the realistic situation, even in the pristine case, as the Majorana oscillations must be accounted for.

In studying MZM splitting in Majorana nanowires in the presence of disorder and wire length, we explicitly focus on the MZM splitting in the first oscillatory lobe beyond TQPT with the minimal MZM splitting, and study its dependence on various system parameters such as disorder strength $\sigma$, wire length $L$, etc.
Other lobes (i.e., at higher Zeeman splitting or lower chemical potential) have larger splitting, and are, therefore, less topological in some crude qualitative sense since MZM splitting implies a breaking of topology due to MZM overlap from the two ends.
This work defines the splitting in the first lobe as the MZM splitting.
Of course, whether this splitting behaves exponentially in system length or not is really the crux of the physics we study in great depth in this work. We obtain the statistical distribution of the splitting as a function of disorder for various wire lengths, and explicitly check whether the splitting is exponentially suppressed.
A key finding is that disorder severely constrains the exponential behavior for a disorder strength smaller than the pristine SC gap, imposing a severe constraint on the meaning of `topology' in finite-length disordered wires since our earlier work indicates the SC gap surviving up to a larger disorder strength comparable to the pristine gap~\cite{pan2020physical,pan2021threeterminal}.
Thus, Majorana nanowires may be more vulnerable to disorder than was believed before based on the numerical simulations of topological indicators.~\cite{dassarma2023spectral,dassarma2023density,pan2024disordered,pikulin2021protocol}

We mention that we ignore any self-energy corrections in the proximity tunneling effect in the main text, where we consider just the pristine proximity effect arising from the leading order tunneling of Cooper pairs from the parent superconductor to the nanowire.~\cite{sau2010nonabelian, sau2010robustness,stanescu2013superconducting} 
The model of bare proximity tunneling is the minimal consistent theory for Majorana nanowires, which is used extensively in theory and simulations in the literature.  Going beyond the minimal theory, one can include the self-energy effect associated with multiple tunneling at the interface by integrating out the fermions, but such calculations are no longer unique and involve additional approximations of how precisely the parent superconductor is integrated out since the full theory including both the parent superconductor and the nanowire is simply intractable theoretically and computationally.  We do provide results in the appendices the situation including multiple tunneling self energy corrections in the proximity effect, and our conclusions from the main text without self-energy effects remain completely valid in the presence of self-energy.  The main effect of self energy is an overall quantitative suppression of the Majorana splitting without any qualitative effect on our conclusion.  Since the actual tunneling strength at the SM-SC interface producing the proximity effect is unknown,  the details of the theoretical modeling for the proximity effect become academic.  The important parameter for comparison with the Majorana splitting is the proximity induced gap, and we discuss in depth the comparison between the gap and the splitting.

The rest of the paper is organized as follows.
In Sec.~\ref{sec:model}, we provide our model and the basic theory.
In Sec.~\ref{sec:results}, we present our results with a detailed discussion.
In Sec.~\ref{sec:summary}, we conclude with a summary and future directions.
{In Appendix~\ref{app:nw_extended}, we provide some extended statistics data for the nanowire model.
In Appendix~\ref{app:coherence_length}, we provide the calculation of the SC coherence length in both pristine and disordered nanowires.
In Appendix~\ref{app:correlation}, we study the correlation between the MZM splitting and the topological invariant. }
In Appendix~\ref{app:bandstructure}, we provide more examples of the band structure for the finite disordered nanowire.
{In Appendix~\ref{app:self_energy}, we show the same type of results in the nanowire model including the self-energy of the superconductor.}
In Appendix~\ref{app:kitaev_chain}, we provide results with similar conclusions for the Kitaev chain model, where the spinless $p$-wave SC is assumed \textit{a priori}.
\section{Model and Theory}\label{sec:model}
\subsection{1D Majorana nanowire model}
We focus on the well-accepted and extensively-used minimal 1D single-band SM-SC Majorana nanowire model with the following Hamiltonian~\cite{lutchyn2010majorana,sau2010generic,sau2010nonabelian,oreg2010helical}:
\begin{equation}\label{eq:hatH}
    \hat{H}=\frac{1}{2}\int dx \Psi^\dagger(x) \left( H_{\text{SM}}+H_{\text{Z}}+H_{\text{SC}}+H_{\text{dis}} \right) \Psi(x),
\end{equation}
where $\Psi(x)=(\psi_\uparrow(x),\psi_\downarrow(x),\psi_\downarrow^\dagger(x),-\psi_\uparrow^\dagger(x))^\intercal$ is the Nambu spinor defined such that the single-particle Hamiltonian of the SM part $H_{\text{SM}}$, Zeeman field part $H_{\text{Z}}$, proximity-induced superconducting part $H_{\text{SC}}$, and disorder part $H_{\text{dis}}$ are given by:
\begin{equation}\label{eq:H}
    \begin{split}
        H_{\text{SM}}&=\left( -\frac{\partial_x^2}{2m^*}-i\alpha \partial_x\sigma_y-\mu \right)\tau_z,\\
        H_{\text{Z}}&=V_Z \sigma_x,\\
        H_{\text{SC}}&=\Delta \tau_x,\\
        H_{\text{dis}}&=V(x)\tau_z 
    \end{split}
\end{equation}
to satisfy the particle-hole symmetry with $\mathcal{P}=\tau_y\sigma_y \mathcal{K}$ where $\mathcal{K}$ is the complex conjugation.
Here, the Pauli matrices $\sigma_i$ and $\tau_i$ act on the spin and particle-hole degrees of freedom, respectively, and $m^*$ is the effective mass of the SM nanowire, $\alpha$ is the spin-orbit coupling strength, $\mu$ is the chemical potential, $V_Z$ is the Zeeman field strength, and $\Delta$ is the induced superconducting gap.
Unless otherwise specified, we choose the parameters $m^*=0.1519 m_e$ ($m_e$ is the electron rest mass), $\alpha=0.5$ eV\AA{}, $\mu=1$ meV, $\Delta=0.2$ meV, which are typical for InAs/Al SM-SC hybrid nanowire systems,~\cite{microsoftquantum2023inasal,aghaee2025interferometric} leading to a topological phase transition at $V_Z= \sqrt{\mu^2+\Delta^2}=1.02$ meV and the minimal superconducting coherence length $\xi_{\min}\approx0.78~\mu$m at $V_Z=0$ (where the SC gap is the largest) in the pristine limit (i.e., $H_{\text{dis}}=0$).
{(See Appendix~\ref{app:coherence_length} for the calculation of the coherence length.)}
The Hamiltonian in {Eq.~\eqref{eq:hatH}} is then numerically diagonalized through the finite difference method on a discretized 1D lattice with a fictitious lattice spacing $a=10$ nm.
We consider a phenomenological non-magnetic disorder potential $V(x)$ in the SM due to random impurities, modeled as a Gaussian random potential with zero mean and standard deviation $\sigma$, i.e., $\expval{V(x)V(x')}= \sigma^2\delta(x-x')a$.
Thus, the discretization scale $a$ should be taken as the disorder correlation length which is a reasonable estimate for the expected screened Coulomb disorder in the nanowire. Here, $\sigma$ is the disorder strength characterizing the random disorder in the sample.
We note that the model defined by Eq.~\eqref{eq:hatH} is equivalent to the modeling used by Microsoft in its recent breakthrough Majorana nanowire experiment including the single-band occupancy aspect.~\cite{microsoftquantum2023inasal}
 We do not, however, make any effort to connect our work quantitatively to the experimental measurements because most of the relevant experimental parameters entering Eq.~\eqref{eq:hatH} are unknown (e.g. effective mass, Land\'e $g$-factor, Rashba spin-orbit coupling, and disorder).

\subsection{Maximal Majorana splitting and gap sizes}
The Majorana splitting energy $E_1$ is the lowest-energy state in the topological phase, arising from the overlap of two MZMs localized near the ends of a wire of length $L$.
It is considered the smoking-gun evidence for the existence of MZMs~\cite{dassarma2016how}, which can be analytically expressed for zero disorder (and $L\gg\xi$)~\cite{cheng2009splitting}
\begin{equation}\label{eq:Es}
    E_1 \propto \Delta \frac{\cos(k_F L+\frac{\pi}{4})}{\sqrt{k_F L}}e^{-\frac{L}{\xi}},
\end{equation}
where $k_F$ is the Fermi wave vector in the SM, and $\xi$ is the SC coherence length.
Therefore, for a fixed nanowire length $L$, the Majorana splitting energy $E_1$ oscillates with the Zeeman field $V_Z$ or the chemical potential $\mu$, as they effectively tune the Fermi wave vector $k_F$.
The oscillation amplitude is exponentially suppressed (for $L\gg\xi$) by the ratio $L/\xi$, reflecting the exponential protection of MZMs in the topological phase.
However, this exponential behavior only holds in the clean or weak-disorder limit ($\sigma\rightarrow0$) and the thermodynamic limit ($L\rightarrow\infty$).
In realistic settings with finite wire length and finite disorder, the MZM splitting energy $E_1$ is no longer described by the above expression in Eq.~\eqref{eq:Es}.
Therefore, we numerically compute the MZM splitting energy by diagonalizing the disordered BdG Hamiltonian from Eq.~\eqref{eq:H} for a finite wire of length $L$ and nonzero disorder strength $\sigma$. Our calculation is exact for specific disorder configurations with the given disorder strength.

As an operational definition, we define the maximal MZM splitting energy $E_s$ as the peak value in the first oscillatory lobe as a function of Zeeman field, occurring after the topological gap has fully closed (see the blue arrow in Fig.~\ref{fig:bandstructure}(a)), i.e.,
\begin{equation}\label{eq:E_s}
    E_s= \max_{V_Z\in \text{first lobe}} E_1(V_Z).
\end{equation}
This requirement---to track the maximal splitting in the first lobe after the initial gap closure---imposes a constraint on the disorder strength: we must remain within the weak disorder regime (see Figs.~\ref{fig:bandstructure},~\ref{fig:bandstructure_L0.6}, and ~\ref{fig:bandstructure_L10} for examples).
If the disorder becomes too strong (as explored in Refs.~\cite{pan2020physical,pan2021threeterminal,dassarma2021disorderinduced,dassarma2023spectral,dassarma2023density,pan2024disordered}), the gap-closing and reopening features can vanish, making it impossible to unambiguously identify the ``first lobe'' and thus define $E_s$.
In fact, the system is no longer in a topological phase for such strong disorder, and the MZMs do not exist.
This is the localized Anderson insulator phase of the nanowire with no topology although some aspects of the MZM phenomenology may be mimicked in the tunnel conductance spectroscopy in short wires.~\cite{pan2020physical}

A concomitant feature of the MZM splitting energy beyond the TQPT is the SC gap size, defined as the energy of the second-lowest state (corresponding to the topological gap or band edge in the thermodynamic continuum limit).
In the pristine limit, the topological gap size is on the order of the proximity-induced superconducting gap.
However, the gap can be significantly suppressed in the presence of disorder, potentially eliminating topological protection.
Thus, the gap size is an additional key metric for characterizing the topological phase.
Noting that the SC gap varies as a function of the Zeeman field, therefore, we define the gap size at the same Zeeman field that maximizes the MZM splitting energy $E_s$ in the first lobe (see the red arrow in Fig.~\ref{fig:bandstructure}(a)), i.e.,
\begin{equation}
    \Delta_s = \left. E_2(V_Z) \right|_{V_Z = \argmax E_1(V_Z)}
\end{equation}

Using these two key metrics (i.e., $E_s$ and $\Delta_s$), we define their ratio as an operational definition of topology in finite disordered nanowires $R = \frac{E_s}{\Delta_s}$. 
The condition $R\ll1$ is necessary for MZMs to be well-defined isolated anyons.
This ratio also characterizes the viability of the putative quantum gate operations in the topological regime, as we want $E_s\ll\Delta_s$ (i.e., $R\ll1$) such that we have a long enough upper bound to remain quantum coherent, and a short enough lower bound to remain adiabatic when manipulating the MZMs.
Note that the ratio $R$ is distinct from the other important dimensionless ratio $\xi/L$ defining exponential protection, but in the limit where they are both $\ll1$, they become equivalent, and topological immunity is well-defined. In contrast to $R$, which can be theoretically obtained by direct diagonalization, there is no direct way to estimate the SC coherence length $\xi$ in the presence of disorder.
We mention here that there are other possible ways to define a unique splitting to gap ratio, e.g., take the minimal gap and the MZM splitting at the same Zeeman field as where the minimal gap occurs beyond the TQPT, or choose the splitting and the gap at some other fixed field, or use the maximum splitting and the maximum gap. As long as the splitting and the gap are chosen at the same $V_Z$ value, our qualitative findings are independent of the specific choice for obtaining $R$, although the quantitative results can differ considerably. We have explicitly verified the qualitative independence of $R$ on the specific choice for its definition. It is extremely important to choose the splitting and the gap at exactly the same Zeeman field because this is what makes sense experimentally---choosing the gap and the splitting at random different field values would produce nonsensical results. We believe that the maximal splitting in the first lobe and the corresponding gap at the same field is the most natural way to define $R$ since large splitting is the main enemy of topology. The experiments will be most strongly affected by the largest splitting right after entering the topological regime, and this leads to our unique choice given by Eq.~\eqref{eq:E_s}.  Since the splitting increases and the gap decreases with increasing field, our choice of the first lobe in defining the ratio provides the optimal definition of the operational topology.  If the system is not topological (because of disorder) even by the definition of our Eq.~\eqref{eq:E_s} as used in this work, it is hopeless to look for topology at higher field values.  This has motivated our choice.

In the following, we generate random disorder potentials $V(x)$ for each disorder strength $\sigma$, and study the disorder-averaged value of the maximal MZM splitting energy $E_s$, the gap size $\Delta_s$ as a function of the wire length $L$ and the disorder strength $\sigma$.
For the ratio $R$, we consider two types of disorder averaging: (1) the ``quenched'' average~\cite{mezard1987spin}
\begin{equation}
    R_1 = \expval{\frac{ E_s }{ \Delta_s }},
\end{equation}
and (2) the ``annealed'' average
\begin{equation}
    R_2=\frac{\expval{E_s}}{\expval{\Delta_s}}
\end{equation}
where $\langle \dots \rangle$ denotes averaging over random disorder realizations.
Of course, each sample would have its specific disorder configuration, which would vary with varying system parameters such as $\mu$ and $V_Z$ because impurities typically move around in samples with varying parameters.
We also provide detailed splitting distributions for fixed disorder strengths.

\section{Results and Discussion}\label{sec:results}
\begin{figure*}[ht]
    \centering
    \includegraphics[width=6.8in]{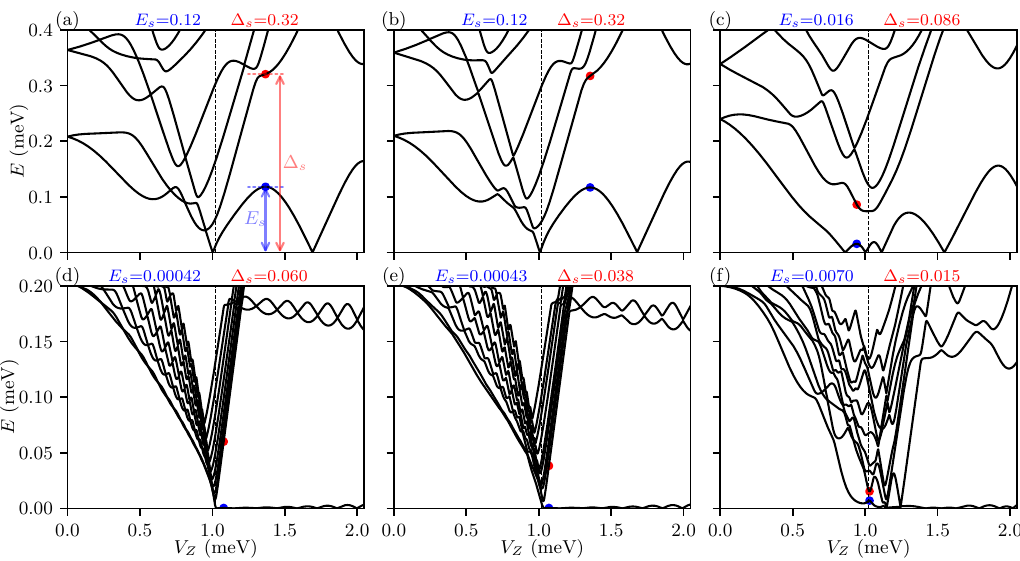}
    \caption{
    Band structure as a function of Zeeman field $V_Z$ for pristine limit $\sigma=0$ (left column), $\sigma=0.1$ meV (middle column), and $\sigma=0.9$ meV (right column) for $L=1~\mu$m (top row) and $L=5~\mu$m (bottom row).
    The maximal MZM splitting energy $E_s$ (in the unit of meV) in the first lobe is indicated by the blue arrow, and the gap size $\Delta_s$ (in the unit of meV) is indicated by the red arrow, with their values at the top of each panel.
    The vertical dashed line indicates the topological phase transition point at $V_Z=1.02$ meV.
    }
    \label{fig:bandstructure}
\end{figure*}
In this section, we present our main results for the maximal MZM splitting energy $E_s$ in the first lobe, the gap size $\Delta_s$, and their ratio $R$, as functions of the disorder strength $\sigma$ and the wire length $L$.
We note that the splitting increases with increasing $V_Z$ (i.e., in the higher lobes); therefore, splitting in the lowest lobe (just above the TQPT) is the smallest possible splitting in the topological phase.~\cite{cheng2009splitting,cheng2010tunneling,dassarma2016how}
This is also where the topological gap is the largest since the gap decreases with increasing $V_Z$ right after reaching the maximum near the TQPT.
In Fig.~\ref{fig:bandstructure}(a), we show the band structure of the nanowire as a function of the Zeeman field $V_Z$ in the pristine limit for $L=1~\mu$m.
We focus on the topological phase ($V_Z>1.02$ meV), and identify the maximal MZM splitting energy in the first lobe, marked by the blue arrow.
At the same Zeeman field, we define the gap size $\Delta_s$ as the energy of the second-lowest state, indicated by the red arrow.
In Fig.~\ref{fig:bandstructure}(b), we increase the disorder strength to $\sigma=0.1$ meV and show a typical realization of the disorder profile $V(x)$ before averaging.
The band structure remains qualitatively similar to the pristine case, indicating topological protection against weak disorder.
However, as we further increase the disorder strength to $\sigma=0.9$ meV in Fig.~\ref{fig:bandstructure}(c), the band structure is significantly modified.
After the initial gap closure, the ``first lobe'' shifts to a smaller Zeeman field ($V_Z \approx 0.9$ meV), accompanied by a substantial reduction in both $E_s$ and $\Delta_s$.
(This disorder-induced shift has sometimes been misleadingly described as a disorder-induced enhancement of topology, it is nothing of that sort, it is just a manifestation of the mesoscopic fluctuations induced by disorder in the system.)

In the bottom row of Fig.~\ref{fig:bandstructure}, we present the band structure for a longer wire with $L=5~\mu$m. We find that the maximal MZM splitting energy $E_s$ in the first lobe is already very small even in the pristine limit, as it is exponentially suppressed by the longer wire length $L$ (see Eq.~\eqref{eq:Es}).
Additionally, the Zeeman field at which $E_s$ is maximized shifts closer to the TQPT point, due to faster Friedel oscillations $\Delta k_F\sim \Delta V_Z \propto \frac{\pi}{L}$.

All band structures shown in Fig.~\ref{fig:bandstructure} correspond to a single disorder realization as a representative example. In the following sections, we will present statistical results based on multiple disorder realizations.
To draw statistical conclusions for disorder-averaged ensembles, we emphasize that there are large fluctuations and the situation for a specific disorder configuration may deviate considerably from the statistical results, particularly for large disorder.
We also discuss the details of the distribution itself below.

\subsection{Maximal Majorana splitting in the first lobe}
We first present the distribution of the maximal MZM splitting energy $E_s$ as a function of disorder strength $\sigma$ for various system sizes $L$ in Fig.~\ref{fig:Es_dist} {(For interested readers, we provide extended data in Appendix~\ref{app:nw_extended} in Fig.~\ref{fig:Es_dist_ext}, and the main messages are already delivered in the main text.)}.
In the short-wire regime ($L\leq 1~\mu$m), shown in Fig.~2(a), $E_s$ exhibits a sharp peak at a finite value in the clean and weakly disordered limit ($\sigma \lesssim 0.5$ meV).
As $\sigma$ increases beyond $\gtrsim 0.5$ meV, the distribution broadens and develops a peak at zero splitting energy, signaling a crossover from ``strong topology immunity'' to ``weak topology immunity'' or ``no immunity'' with increasing disorder.
In contrast, for longer wires, the distribution of $E_s$ becomes less sensitive to $\sigma$ and remains sharply peaked near zero splitting energy across all disorder strengths. This insensitivity arises because in the long-wire limit, $E_s$ is already exponentially suppressed, making it effectively indistinguishable from zero even in the presence of disorder.
Note that it is possible to be `lucky', even for strong disorder in a particular configuration because the splitting may vanish because of the broad distribution in the splitting, and such rare  `lucky' event may lead to topologically `patchy' regime~\cite{pan2020physical,dassarma2021disorderinduced,dassarma2023spectral,dassarma2023density,pan2024disordered} since the slightest change in the system parameters (e.g., Zeeman splitting, chemical potential) is likely to suppress this fine-tuned `luck', leading to a large splitting because of mesoscopic fluctuations apparent in the broad splitting distribution for large disorder.
Thus, even if one gets `lucky' in achieving topology arising from mesoscopic fluctuation, this `topological luck' is highly fine-tuned and would disappear with the slightest change in system parameters or even in the disorder configuration even if the disorder strength does not change.  This is not robust topology and obviously there is little protection in such a fine-tuned topology by ``luck'' scenario.

This difference in the distribution of the maximal MZM splitting energy $E_s$ between the short and long wires can be more clearly seen by directly examining the disorder-averaged maximal MZM splitting energy $\expval{E_s}$ as a function of disorder strength $\sigma$ in Fig.~\ref{fig:Es_vs_sigma}(a) and (b).
For very short wires ($L\sim 0.6~\mu$m), $\expval{E_s}$ increases with the disorder strength $\sigma$, which is a typical finite size effect as shown in Fig.~\ref{fig:bandstructure_L0.6}.
As the wire length increases to around 1 $\mu$m, $\expval{E_s}$ reaches its maximum in the pristine limit and decreases as $\sigma$ increases.
In the long-wire limit ($L\gtrsim 3~\mu$m), $\expval{E_s}$ is already very small even in the pristine wire limit, due to exponential topological protection from the two MZMs at two ends of the wire being separated by more than five coherence lengths ($\xi\approx 0.78~\mu$m).
In this regime, $\expval{E_s}$ increases with $\sigma$ and eventually saturates at a finite value as disorder becomes stronger.

We also present the disorder-averaged maximal MZM splitting energy $\expval{E_s}$ as a function of the wire length $L$ for various disorder strengths $\sigma$ in Fig.~\ref{fig:Es_vs_L}.
In the pristine limit, we find an initial nonmonotonic increase in $\expval{E_s}$ as the wire length $L$ increases in the very short-wire limit ($L\lesssim 1~\mu$m), due to the finite size effect (and this regime is nontopological any way since the wire length is comparable to the coherence length).
Beyond this regime $L\gtrsim 1~\mu$m, $\expval{E_s}$ begins to decay exponentially with $L$.
The exponential decay of $\expval{E_s}$ is quantitatively consistent with the theoretical prediction in Eq.~\eqref{eq:Es}, where $E_s = \max(E_1) \propto e^{-\frac{L}{\xi}}L^{-\frac{1}{2}}\sim e^{-\frac{L}{\xi}}$ at leading order. This is the topological regime with `exponential protection'.
The lower-left inset of Fig.~\ref{fig:Es_vs_L} shows the best-fit curve (red dashed line), yielding a coherence length $\xi_s=0.77(2)~\mu$m, which closely matches the minimal coherence length $\xi_{\min} \approx 0.78~\mu$m estimated from the wavefunction amplitude $\abs{\psi(x)}\sim e^{-x/\xi}$, demonstrating the exponential topological protection of MZMs in the pristine long-wire limit.

However, as the disorder strength $\sigma$ increases, the maximal MZM splitting energy $E_s$ begins to deviate from the exponential decay, and instead saturates at a finite value as the wire length $L$ increases.
In this disordered regime, $E_s$ is not exponential in $L$.
In the upper right inset of Fig.~\ref{fig:Es_vs_L}, we fit the disorder-averaged splitting energy $\expval{E_s}$ as a function of $L$ in the disordered case to a power-law decay for the large wire length $L\gtrsim 5~\mu$m, i.e., $\expval{E_s} = \alpha L^{-\eta}$, where the best fit yields $\alpha=0.010(5)$ meV and $\eta = 1.7(2)$.
This breakdown of exponential decay in the disordered case happens around $\sigma \gtrsim 0.2$ meV, comparable to the pristine superconducting gap $\Delta=0.2$ meV.
Thus, the system is trivial for disorder comparable to the pristine zero-field gap, no matter how long the wire might be.
This means that there is nothing to be gained by increasing the wire length unless the disorder is decreased simultaneously in such a situation.  We believe that the recent Microsoft experiment is in this regime. ~\cite{microsoftquantum2023inasal}
{For more examples of the exponential to power-law decay crossover, see Fig.~\ref{fig:Es_vs_L_ext} in Appendix~\ref{app:nw_extended}.}

We also study the Kitaev chain, and observe similar crossover, as discussed in Appendix~\ref{app:kitaev_chain}.
However, the topology based on MZM splitting in the Kitaev chain is suppressed for effectively smaller disorder than in the nanowire. However, a direct quantitative comparison is inapplicable since the models are different.~\cite{pan2023majorana}
The Kitaev chain seems to be quantitatively more sensitive to disorder than the SM-SC nanowire, although their qualitative behavior is similar.
{For similar results in the nanowire model considering the self-energy of the superconductor~\cite{sau2010robustness}, see Appendix~\ref{app:self_energy}.}

\begin{figure}[ht]
    \centering
    \includegraphics[width=3.4in]{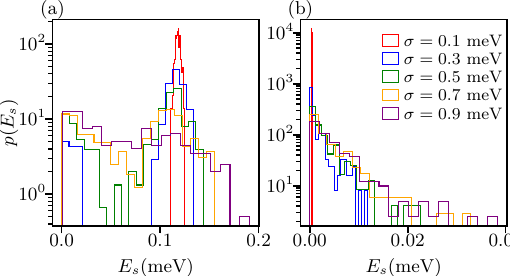}
    \caption{
    The distribution of the maximal MZM splitting energy $E_s$ in the first lobe {for different disorder strength $\sigma$} for (a) $L=1~\mu$m and (b) $L=5~\mu$m. 
    }
    \label{fig:Es_dist}
\end{figure}
\begin{figure}[ht]
    \centering
    \includegraphics[width=3.4in]{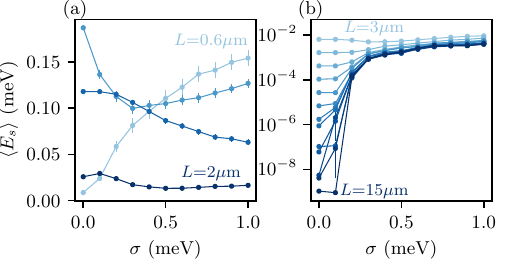}
    \caption{
    The disorder-averaged maximal MZM splitting energy $\expval{E_s}$ in the first lobe as a function of disorder strength $\sigma$ for (a) short wires ($L=0.6,0.8,1,2~\mu$m from light to dark blue) and (b) long wires ($L=3,4,5,\dots,15~\mu$m from light to dark blue).
        }
    \label{fig:Es_vs_sigma}
\end{figure}

\begin{figure}[ht]
    \centering
    \includegraphics[width=3.4in]{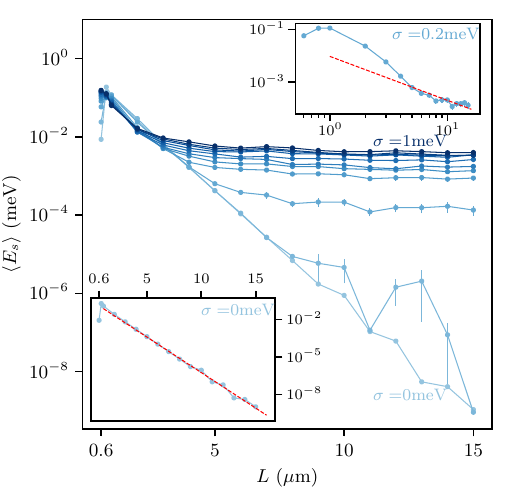}
    \caption{
        The disorder-averaged maximal MZM splitting energy $\expval{E_s}$ in the first lobe as a function of the wire length $L$ for different disorder strengths $\sigma$ (from light to dark blue: $\sigma=0,0.1, 0.2,\dots,1$ meV).
        Lower-left inset: The exponential fit of $\expval{E_s}$ in the pristine limit $\sigma=0$ for $L\ge5~\mu$m with $ \expval{E_s}= A_{s}\exp(-L/\xi_s)$, where the best fit $A_{s}=0.27(8)$ meV and $\xi_s=0.77(2)~\mu$m.
        Upper-right inset: The power-law fit of $\expval{E_s}$ for $L\ge5~\mu$m at the crossover point $\sigma=0.2$ meV in the log-log plot with $\expval{E_s} = \alpha_{s} L^{-\eta_{s}}$, where the best fit $\alpha_{s}=0.010(5)$ meV and $\eta_{s} = 1.7(2)$.
        }
    \label{fig:Es_vs_L}
\end{figure}

\subsection{Gap size}

\begin{figure}[ht]
    \centering
    \includegraphics[width=3.4in]{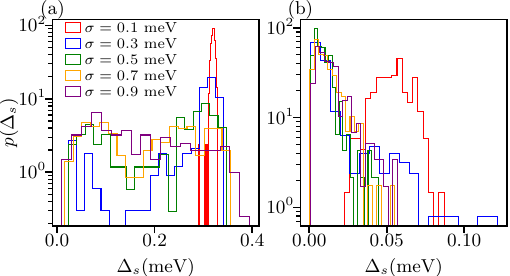}
    \caption{
        The distribution of the gap size $\Delta_s$ {for different disorder strength $\sigma$} for (a) short wires ($L=1~\mu$m) and (b) long wires ($L=5~\mu$m) .
    }
    \label{fig:gap_dist}
\end{figure}

Another key metric for characterizing the topological phase is the gap size $\Delta_s$ at the same Zeeman field as the maximal MZM splitting energy $E_s$.
In Fig.~\ref{fig:gap_dist}(a), we show the distribution of $\Delta_s$ as a function of disorder strength $\sigma$ for a short wire ($L=1~\mu$m), and in Fig.~\ref{fig:gap_dist}(b), for a long wire ($L=5~\mu$m).
These distributions provide a measure of the mesoscopic fluctuations in the SC gap in realistic nanowires even if they have nominally the same disorder (or equivalently mobility or conductance).
Here, we find that the gap size $\Delta_s$ is generally peaked at a finite value in the zero and very weak disorder limit ($\sigma \lesssim 0.5$ meV). At the same time, the distribution broadens and develops a peak at zero gap for $\sigma \gtrsim 0.5$ meV, similar to the maximal MZM splitting energy $E_s$.
The magnitude of gap size $\Delta_s$ is also strongly suppressed in the long wire limit ($L=5~\mu$m) as shown in Fig.~\ref{fig:gap_dist}(b).
{We also present the gap distribution at other system sizes, and another way of extracting the gap as the minimal energy in the continuum spectrum in the entire topological regime after the gap reopens in Fig.~\ref{fig:gap_dist_ext} in Appendix~\ref{app:nw_extended}.}

\begin{figure}[ht]
    \centering
    \includegraphics[width=3.4in]{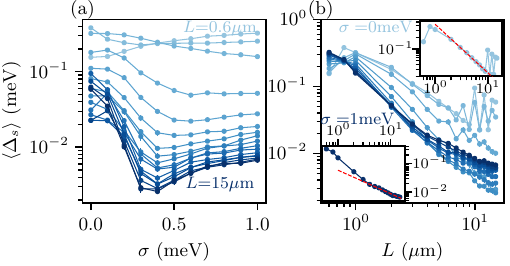}
    \caption{
    The disorder-averaged gap size $\expval{\Delta_s}$ as a function of (a) disorder strength $\sigma$ for different wire lengths $L$ (from light to dark blue: $L=0.6,0.8,1,2,3,\dots,15~\mu$m) and (b) wire length $L$ for different disorder strengths $\sigma$ (from light to dark blue: $\sigma=0,0.1,0.2,\dots,1$ meV).
    Lower-left inset: The power-law fit of $\expval{\Delta_s}$ for $L\ge5~\mu$m at $\sigma=1$ meV, 
    where the best fit $\expval{\Delta_s} = \alpha_{\Delta} L^{-\eta_{\Delta}}$ gives $\alpha_\Delta= 0.058(3)$ and $\eta_\Delta = 0.82(3)$. 
    Upper-right inset: The power-law fit of $\expval{\Delta_s}$ for $2\le L \le 10 ~\mu$m in the pristine limit, where the best fit gives $\alpha_\Delta= 0.43(2)$ and $\eta_\Delta = 1.25(3)$.
    }
    \label{fig:gap_scaling}
\end{figure}

Similarly, we present the disorder-averaged gap size $\expval{\Delta_s}$ as a function of the disorder strength $\sigma$ in Fig.~\ref{fig:gap_scaling}(a) and as a function of the wire length $L$ in Fig.~\ref{fig:gap_scaling}(b).
In the very short wire limit ($L\lesssim 1~\mu$m), the gap size $\expval{\Delta_s}$ increases with the disorder strength $\sigma$, which again is due to the finite size effect, where the ``gap closing'' and ``gap reopening'' feature is not sharp in the band structure (see Fig.~\ref{fig:bandstructure_L0.6}).
This is, of course, mostly nontopological behavior in short wires.
In a longer wire regime ($1\lesssim L\lesssim 5~\mu$m), the gap size $\expval{\Delta_s}$ drastically decreases as the disorder strength $\sigma$ increases, eventually saturating at a finite value for $\sigma\gtrsim 0.5$ meV.
In the long wire limit ($L\gtrsim 5~\mu$m), this behavior results in a nonmonotonic trend where $\Delta_s$ first drops and then saturates, with a minimum occurring around $\sigma \approx 0.3$ meV.

In Fig.~\ref{fig:gap_scaling}(b), we present the disorder-averaged gap size $\expval{\Delta_s}$ as a function of the wire length $L$ for various disorder strengths $\sigma$.
We find that $\expval{\Delta_s}$ decreases algebraically with the wire length $L$ in both the pristine limit and the disordered case, as shown in the upper right and lower left insets, respectively.
In the pristine limit, this behavior arises because the Zeeman field $V_Z$ at which the maximal MZM splitting $E_s$ occurs lies at an energy offset $\Delta V_Z \equiv V_Z-V_{Zc}=V_Z-\sqrt{\mu^2+\Delta^2}$ from the TQPT point, and this offset scales inversely with $L$.
This scaling follows from: (a) $\Delta k_F\sim 1/L$ due to the $\cos(k_F L +\pi/4)$ Friedel oscillation factor in Eq.~\eqref{eq:Es}, and (b) $\Delta k_F \sim \Delta V_Z$ for $V_Z>\mu$ in the topological phase.
\footnote{In the SM (i.e., $\Delta=0$) for $V_Z>\mu$, there is a unique Fermi wave vector $k_F=\left[ 2\left( m^2\alpha^2+m\mu + m\sqrt{V_Z^2+m^2\alpha^4+2m\alpha^2\mu} \right) \right]^{1/2}=2\sqrt{m(m\alpha^2+\mu)}+\frac{1}{2}\mu\sqrt{\frac{m}{(m\alpha^2+\mu)^3}}\Delta V_Z$, where $\Delta V_Z=V_Z-\mu$.}
Near the TQPT, the gap size $\Delta_s$ is governed by the band edge at $k = 0$ in the continuum limit, giving $\Delta_s=E_1(k=0)=V_Z-\sqrt{\mu^2+\Delta^2}$.
Therefore, the gap size $\Delta_s$ scales inversely with $L$ in the pristine limit, which is qualitatively consistent with the power-law fit in the upper right inset of Fig.~\ref{fig:gap_scaling}(b) with $\expval{\Delta_s} = \alpha_\Delta L^{-\eta_\Delta}$, with best-fit parameters $\alpha_\Delta= 0.43(2)$ meV and $\eta_\Delta = 1.25(3)$.

In the disordered case, we find numerically that $\expval{\Delta_s}$ continues to exhibit power-law decay with $L$, as shown in the lower-left inset of Fig.~\ref{fig:gap_scaling}(b), with best-fit values $\alpha_\Delta= 0.058(3)$ meV and $\eta_\Delta = 0.82(3)$.

\subsection{Ratio as an operational definition of topology}
Therefore, we define an operational definition of topology in finite disordered nanowires as the ratio of the maximal MZM splitting energy $E_s$ to the gap size $\Delta_s$, both at the same Zeeman field.
In the pristine limit, this ratio asymptotically approaches zero due to the exponential decay of the MZM splitting energy $E_s$ as a function of the wire length $L$.
In contrast, in the presence of disorder, the ratio can remain finite even in the long-wire limit, since both $E_s$ and $\Delta_s$ decay algebraically with $L$.

\begin{figure}[ht]
    \centering
    \includegraphics[width=3.4in]{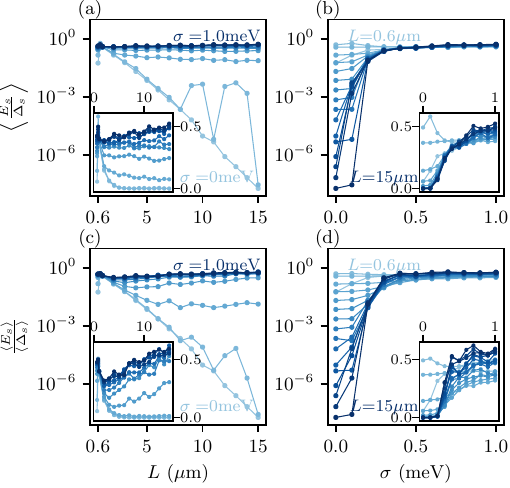}
    \caption{
        The quenched average $R_1=\expval{\frac{E_s}{\Delta_s}}$ (top row) and the annealed average $R_2=\frac{\expval{E_s}}{\expval{\Delta_s}}$ (bottom row) as a function of the wire length $L$ for different disorder strength $\sigma$ (from light to dark blue: $\sigma=0,0.1,0.2,\dots,1$ meV) in the left column (a,c), and as a function of the disorder strength $\sigma$ for different wire lengths $L$ (from light to dark blue: $L=0.6,0.8,1,2,3,\dots,15~\mu$m) in the right column (b,d).
        Insets show the same data in the linear scale.
    }
    \label{fig:ratio}
\end{figure}

In Fig.~\ref{fig:ratio}(a, c), we show the quenched and annealed averages of this ratio as a function of $L$, demonstrating the transition from the exponential decay to the algebraic decay as the disorder strength $\sigma$ increases.
Both averaging methods yield qualitatively similar results.
In practice, this ratio can distinguish between weak and strong disorder regimes for sufficiently long wires, as it can differ by several orders of magnitude between the two limits, as shown in Fig.~\ref{fig:ratio}(b, d). Thus, the ratio serves as an operational definition of the topology in the finite disordered nanowire.
{We present more examples of the ratio in Fig.~\ref{fig:ratio_ext} at other $V_Z$ in Appendix~\ref{app:nw_extended}, which show similar qualitative behavior.}

\subsection{Electron localization length in the SM}
\begin{figure}[ht]
    \centering
    \includegraphics[width=3.4in]{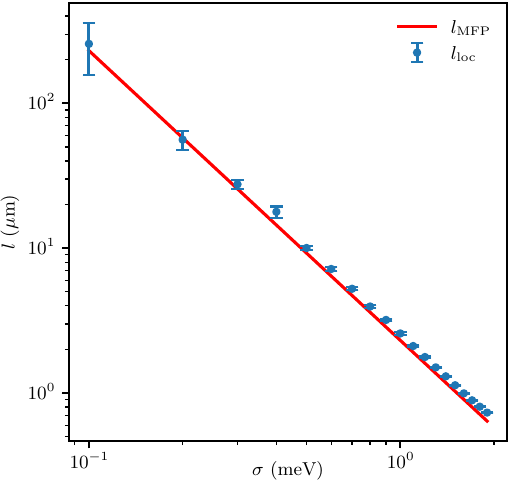}
    \caption{Localization length $l_{\text{loc}}$ in Eq.~\eqref{eq:l_loc}  (blue dots) and the mean free path $l_{\text{MFP}}$ in Eq.~\eqref{eq:l_MFP} (red line) as a function of disorder strength $\sigma$.}
    \label{fig:localization_length}
\end{figure}

Finally, we calculate the electron localization length $l_{\text{loc}}$ in the SM as a function of the disorder strength $\sigma$ as shown in Fig.~\ref{fig:localization_length}.
Note that this differs from the Majorana localization length, where the latter is quantitatively closer to the SC coherence length $\xi\approx 0.78~\mu$m here.
The maximum possible value of the electron localization length is $L$, the wire length, but for finite disorder the localization length becomes $<L$, and is the relevant length to be compared with the coherence length $\xi$ to figure out topology.
Here, we consider a bare SM at zero Zeeman field without proximity superconductivity, and use previous parameters in the SM Hamiltonian in Eq.~\eqref{eq:H}.
We compute the electron localization length $l_{\text{loc}}$ from the inverse of the smallest-magnitude negative Lyapunov exponent $\gamma_{\min}$ of the product of all transfer matrices, namely,
\begin{equation}\label{eq:l_loc}
    l_{\text{loc}}=\frac{a}{\gamma_{\min}},
\end{equation}
where
\begin{equation}
    \gamma_{\min} = \min_{\gamma_i < 0} |\gamma_i|, \quad \gamma_i = \lim_{N\to \infty} \frac{1}{N} \ln \left( \prod_{i=1}^N T_i \right),
\end{equation}
Here, $T_i$ is the transfer matrix at site $i$, defined as 
\begin{widetext}
    \begin{equation}
        \begin{pmatrix} 
    \gamma_{i+1,\uparrow} \\ 
    \gamma_{i+1,\downarrow} \\ 
    \gamma_{i,\uparrow} \\ 
    \gamma_{i,\downarrow} 
    \end{pmatrix} = 
    \begin{pmatrix}
    \frac{\alpha _R V_Z+2 t^2-\mu  t}{\alpha _R^2+t^2} & \frac{-\mu  \alpha _R+2 t
    \alpha _R+t V_Z}{\alpha _R^2+t^2} & \frac{\alpha _R^2-t^2}{\alpha _R^2+t^2} &
    -\frac{2 t \alpha _R}{\alpha _R^2+t^2} \\
    \frac{\mu  \alpha _R-2 t \alpha _R+t V_Z}{\alpha _R^2+t^2} & \frac{-\alpha _R
    V_Z+2 t^2-\mu  t}{\alpha _R^2+t^2} & \frac{2 t \alpha _R}{\alpha _R^2+t^2} &
    \frac{\alpha _R^2-t^2}{\alpha _R^2+t^2} \\
    1 & 0 & 0 & 0 \\
    0 & 1 & 0 & 0 \\
    \end{pmatrix}
    \begin{pmatrix}
    \gamma_{i,\uparrow} \\
    \gamma_{i,\downarrow} \\
    \gamma_{i-1,\uparrow} \\
    \gamma_{i-1,\downarrow}
    \end{pmatrix} ,
\end{equation}
\end{widetext}
where $\gamma_{i,\uparrow/\downarrow}$ is one of the Majorana operators at site $i$ for the spin up and down,  $\alpha_R = \frac{\alpha}{2a}$ is the effective Rashba spin-orbit coupling strength, and $t = \frac{1}{2m^* a^2}$ is the effective hopping amplitude in the discretized lattice with lattice constant $a=10$ nm.

We randomize $N=10^6$ different disorder realizations for each disorder strength $\sigma$, and compute the localization length $l_{\text{loc}}$ as shown in the blue markers in Fig.~\ref{fig:localization_length}.
The localization length $l_{\text{loc}}$ diverges at $\sigma=0$ in the pristine limit, and algebraically decays with the disorder strength $\sigma$ with the exponent being approximately $-2$. 

This behavior can be intuitively understood by estimating the mean free path (MFP) in the 1D SM, which can be modeled as
\begin{equation}\label{eq:l_MFP}
    l_{\text{MFP}} \approx v_F \tau \approx  \frac{v_F^2}{\sigma^2 a},
\end{equation}
where $v_F$ is the Fermi velocity at zero Zeeman field, and the last approximation comes from Fermi's golden rule.
Within the same parameters, we estimate the MFP $l_{\text{MFP}}$ as a function of disorder strength $\sigma$ as shown by the red line in Fig.~\ref{fig:localization_length}, showing a quantitative agreement with the localization length $l_{\text{loc}}$.
This is expected since in 1D, the mean free path is the localization length.

Finally, with the estimate of the MFP $l_{\text{MFP}}$/electron localization length $l_{\text{loc}}$, and the SC coherence length/Majorana localization length $\xi\approx 0.78~\mu$m, we want to ensure that $2l_{\text{loc}}>\xi$ to satisfy the condition for topology as discussed in Ref.~\onlinecite{brouwer2011probability} such that the average of the MZM splitting energy remains exponentially small.

However, there are some differences between our work and the conclusion in Ref.~\onlinecite{brouwer2011probability}.
First, Ref.~\onlinecite{brouwer2011probability} is based on the assumption of the thermodynamic limit, while our work focuses on the realistic short wire limit, which is the experimentally relevant situation.
Second, using the estimated MFP in Eq.~\eqref{eq:l_MFP}, we can find that the critical value before the exponential decay regime breaks down is $\sigma\approx2.4$ meV, which is much larger than the threshold $\sigma\approx 0.2$ meV we find in our numerical analysis. Therefore, we are studying in the regime within the topological immunity in Ref.~\onlinecite{brouwer2011probability}, where their conclusion puts an upper bound.
Finally, there is a technical difference in the definition of the MZM splitting energy $E_s$. Here, we define $E_s$ as the \textit{maximal} splitting energy in the first lobe, while Ref.~\onlinecite{brouwer2011probability} considered all possible splitting energies with the same pristine wire parameters (ignoring the role of Majorana oscillations producing the lobes in the splitting energy).
The discrepancy in the definition of $E_s$ can lead to a shift in the crossover point, since in our case, the peak of the MZM splitting energy $E_s$ can occur at different Zeeman fields $V_Z$ across different disorder realizations, which affects $k_F$ and $\xi$.

\section{Summary and Conclusion}\label{sec:summary}

We theoretically study in this work the dependence of Majorana (zero mode) splitting energy on various physical parameters such as disorder and wire length in SM-SC hybrid Majorana nanowire systems, focusing approximately on the experimentally relevant InAs/Al hybrid platform. Since the MZM splitting is suppressed exponentially $\sim \exp(-L/\xi)$, as the `exponential signature' of topological immunity with $L$, $\xi$ being the effective length of the wire (or the effective electron mean free path, whichever is shorter), an understanding of the behavior of the splitting in realistic disordered finite-length nanowires is important both from fundamental and from technological perspectives. In particular, topology necessitates $L\gg\xi$ and MZM splitting much less than the topological SC gap, with disorder strongly affecting all of these relevant parameters, as our extensive analysis establishes. We calculate the MZM splitting distribution as well as the effective $\xi$ by using direct numerical diagonalization of the applicable BdG equation in realistic nanowires in the presence of disorder. We show that the exponential dependence of the topological immunity is strongly and nontrivially affected by disorder, and for disorder of the order of the SC gap, topological immunity is suppressed, even if the system is putatively in the corresponding pristine topological phase. Our work connects the MZM splitting directly with topology, which is, in some sense, the essence of `topological immunity' as was already emphasized by Kitaev in his early work on MZMs in pristine long 1D wires, but we show that the situation is drastically different in disordered wires of finite length. In particular, disorder of the order of the pristine gap would completely suppress all exponential protections.
We emphasize that for any finite disorder there is a possibility that, although most configurations are topological, a specific disorder configuration in a given sample has no topology since the MZM splitting has a distribution, which broadens with increasing disorder. 
Thus, for larger disorder, most configurations would be nontopological, but it is possible, because of mesoscopic fluctuations associated with the broad distribution of the splitting, that a fine-tuned particular configuration happens to be topological, but such topology will be very fragile, disappearing with small changes in the parameters.

Although our work is conceptual and theoretical, focusing on the fundamental role disorder and wire length play in defining topology, it has obvious implications for experiments since we show that disorder and finite wire length could severely suppress the topological phase and the exponential MZM protection. For example, the active experimental research on InAs/Al SM-SC platform manifests a pristine zero-field gap $\sim 0.15$ meV, with a claimed topological gap of 0.02--0.04 meV, in nanowires of $L\sim3$ microns.
Our work therefore suggests that the disorder strength must be smaller than 0.15 meV for any manifestation of exponential protection.
This threshold depends on detailed parameters such as the fictitious lattice constant $a$ and, therefore, may differ in experimental samples; however, there is a disorder threshold comparable to the proximity gap.
A disorder strength (or equivalently, impurity broadening) of 0.15 meV corresponds to an electron mobility of $\sim$ 200000 cm$^2$/V.s. The current experimental systems have mobilities in the range of 50000--100000 cm$^2$/V.s.,~\cite{microsoftquantum2023inasal,aghaee2025interferometric} which may be consistent with the reported fragmented and patchy topological phase with very small ($\sim$ 0.03 meV) topological gap, as also emphasized in the recent literature.~\cite{dassarma2023density,dassarma2023spectral} 
It appears that the experimental systems may be a factor of 2--3 larger than the disorder constraint our work imposes, perhaps explaining why the current gap and topological regimes are so small and patchy since we find very broad distribution in the presence of disorder which could produce patchy and fragile topological regimes even for relatively strong disorder.
However, we add that the experimental systems are complex, and the extent to which our minimal Majorana nanowire model applies to these samples is unknown. So, our comments here on the experiments should be taken as qualitative or at best semi-quantitative, and not decisive.

The fact that determining topology of a Majorana nanowire, particularly in the realistic SM-SC hybrid platforms used experimentally, is a difficult challenge has been strongly emphasized by our group for more than 10 years in many publications, with the early work focusing on the non-reliability of using ZBCP by itself as a signature of topological MZMs,~\cite{dassarma2016how,liu2017andreev,liu2018distinguishing,chiu2019fractional,chiu2017conductance,huang2018metamorphosis,pan2020generic,pan2022random,pan2020physical,dassarma2021disorderinduced,ahn2021estimating,pan2021quantized,pan2022ondemand} and the more recent work showing that even the combination of local and nonlocal tunneling may not always decisively determine topology in finite disordered wires and additional considerations may be necessary~\cite{dassarma2023density,dassarma2023spectral,pan2024disordered,taylor2024machine,taylor2025vision}. The possibility that a particular putative topological signature arises from trivial causes cannot be absolutely ruled out based only on transport measurements as emphasized by us has also been recently discussed by other groups~\cite{day2025identifying}. The problem would go away if the predicted Majorana quantization is routinely and robustly seen in many samples, but this is an unlikely prospect, and even then, the non-Abelian topological nature of MZMs must be established by some interference/braiding measurement. This can be seen from the history of the even-denominator fractional quantum Hall states in 2D systems, where, almost 40 years after the experimental discovery of the 5/2 FQHE state~\cite{willett1987observation} and 20 years after the proposed topological qubit using the 5/2 state~\cite{dassarma2005topologically}, their theoretically predicted non-Abelian topological properties are not well-established in spite of some recent progress.~\cite{willett2023interference} Our current work should be viewed from this perspective of the severe difficulties in defining topology for finite disordered systems, since mathematically quantum topology is a pristine property in the thermodynamic limit. 
The topology in this case arises from an intrinsic quantum degeneracy of the ground state, which arises directly from the exponentially small MZM splitting--- if the splitting is algebraic, there is no quantum degeneracy, and no topology.
Our careful and detailed study of MZM splitting as a function of disorder and system size shows why the problem is challenging, and what the issues are in achieving progress in the field.~\cite{hegde2016majorana,penaranda2018quantifying,avila2020majorana,cayao2017majorana,zyuzin2013correlations}

One physical way to understand the detrimental role of the Majorana splitting from the MZM overlap is to realize that the SC gap $\Delta$ must be much larger than the splitting energy $E_s$ to avoid qubit decoherence, i.e., large $E_s$ violates the fundamental condition $E_s\ll\Delta$ essential for topological protection.~\cite{cheng2011nonadiabatic,cheng2012topological} For example, any gate operation must be carried out over a timescale $\tau$ fast enough such that $1/\tau \gg E_s$, but it must also be slow enough so that $1/\tau\ll\Delta$; thus the essential constraint for decoherence-free topological operations is: $\Delta\gg 1/\tau\gg E_s$. This condition can only be satisfied if $\Delta\gg E_s$, i.e., $E_s$ is small. If the system is in the exponential regime, $L\gg\xi$, then $E_s \sim \Delta \exp(-L/\xi)$, and therefore, $\Delta\gg E_s$ is then automatically satisfied. In general, $\xi \sim 1/\Delta$, and a small topological gap implies a large $\xi$, inhibiting the $L\gg\xi$ condition. Of course, a small $L$ also would violate the $L\gg\xi$ condition. Since disorder generally suppresses the topological gap, we need low disorder and large $L$ to obtain topological immunity or topological protection.
Note that in the presence of strong disorder $L$ should be replaced by $l_{\text{loc}}$ (when $l_{\text{loc}}<L$ is satisfied) in these considerations, further suppressing the $L$ or $l_{\text{loc}}\gg \xi$ necessary condition.  We have shown explicitly that the Majorana splitting is nonexponential in $L$ for disorder of the order of or larger than the induced SC gap.
Since the recent experimental wire length for the Microsoft samples is 3 microns~\cite{microsoftquantum2023inasal,aghaee2025interferometric}, the putative coherence length is then roughly $\xi\sim 0.3$ microns, which very approximately coincides with the estimated coherence length for the Al/InAs SM-SC platform. We cannot, however, be sure that this is what is transpiring in the Microsoft sample since their $X$ gate data are noisy and the possibility that the switching time in the experiment is not related to the MZM splitting arising instead purely from some intrinsic noise in the device (e.g., typical charge noise induced switching in semiconductors) cannot be ruled out. More experimental work should resolve this important question.~\cite{nayak2025realizing}
Finally, we note that an early experiment~\cite{albrecht2016exponential} claimed the observation of exponential protection in disordered InAs/Al samples, but careful theoretical analyses established that the relevant physics here is disorder and multisubband effects, and not arising from any exponential behavior of Majorana splitting.~\cite{chiu2017conductance,lai2021theory}.
To our knowledge, such an exponential protection has never been claimed in any other experimental publication.

We conclude by commenting on the MZM splitting distribution in the presence of disorder.  
In the thermodynamic limit, this distribution has been conjectured to be log-normal~\cite{brouwer2011probability}, and for long wires and weak disorder this indeed seems to be the case (see, e.g., Fig.~\ref{fig:Es_dist}(b) and other similar figures in the appendix). 
This thermodynamic log normal distribution is, however, of academic interest since in realistic wires the distribution is in fact peaked at a finite energy (e.g. Fig.~\ref{fig:Es_dist}(a)) and can even be bimodal. 
The same is true for the distribution of the gap, see Fig.~\ref{fig:gap_dist} and several examples in the appendix.
The physics is complex for finite wires with strong disorder where random matrix theories applicable for infinite systems do not apply, and the Majorana splitting is by no means exponentially small.  
Another qualitative difference is that in the thermodynamic limit, there are no MZM oscillations, and the problem simplifies considerably.
Finally, our statistical distribution focuses on the maximal MZM splitting after the gap reopens which is more relevant experimentally, compared with the distribution of all possible splittings at all Zeeman fields.

\section{Acknowledgements}
H.P. is supported by US-ONR grant No.~N00014-23-1-2357.
This work is supported by the Laboratory for Physical Sciences through the Condensed Matter Theory Center at the University of Maryland.

\bibliography{Paper_splitting}
\appendix
\section{Extended data for statistics in the nanowire model}\label{app:nw_extended}
\begin{figure*}[ht]
    \centering
    \includegraphics[width=6.8in]{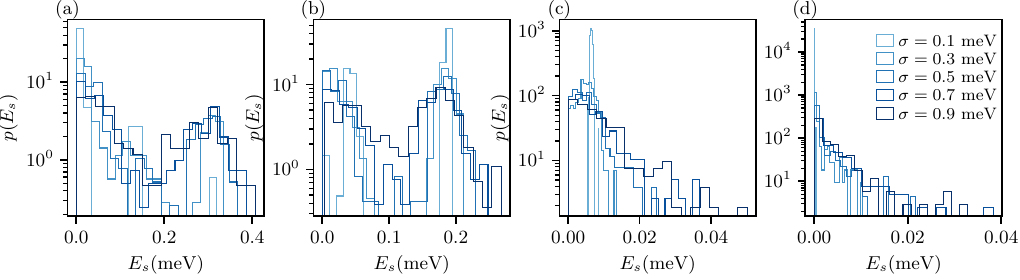}
    \caption{
        The distribution of maximal MZM splitting energy $E_s$ in the first lobe as a function of disorder strength $\sigma$ for (a) $L=0.6~\mu$m, (b) $L=0.8~\mu$m, (c) $L=3~\mu$m, and (d) $L=10~\mu$m.
    }
    \label{fig:Es_dist_ext}
\end{figure*}

In this section, we present some extended data for the nanowire model in addition to the main text.
Since the appendix figures present the same types of results as in the main text but for different system parameters (e.g., wire length, disorder strength, and magnetic field), the figure captions and descriptions are necessarily similar, differing primarily in the specific parameter values used.
In Fig.~\ref{fig:Es_dist_ext}, we show distribution of maximal MZM splitting energy $E_s$ in the first lobe as a function of disorder strength $\sigma$ for $L$ = 0.6, 0.8, 3, 10 $\mu$m.

\begin{figure*}[ht]
    \centering
    \includegraphics[width=6.8in]{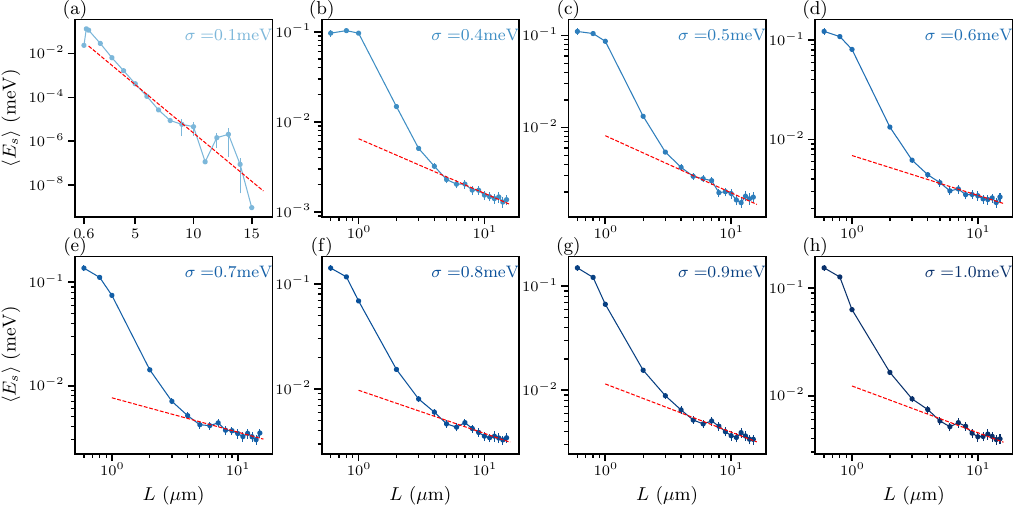}
    \caption{
        The exponential to power-law crossover of the disorder-averaged maximal MZM splitting energy $\expval{E_s}$ in the first lobe as a function of the wire length $L$ for different disorder strengths $\sigma$=0.1, 0.4, 0.5, 0.6, 0.7, 0.8, 0.9, 1 meV in (a-h), respectively. Here, $\sigma =0.1$ meV shows the exponential decay $\xi_s=0.93~\mu$m, while for $\sigma=0.4$ to 1 meV, they show the power-law decay with $\eta_s =$ 0.60, 0.62, 0.40, 0.33, 0.41, 0.46, and 0.43 respectively.
    }
    \label{fig:Es_vs_L_ext}
\end{figure*}

In Fig.~\ref{fig:Es_vs_L_ext}, we show fitting from exponential to power-law decay of the disorder-averaged maximal MZM splitting energy.

\begin{figure}[ht]
    \centering
    \includegraphics[width=3.4in]{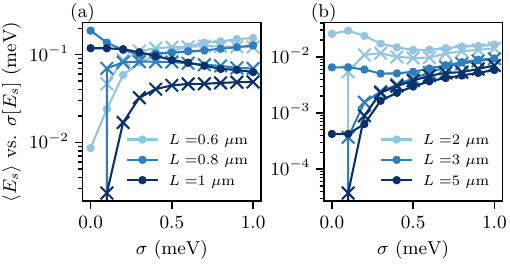}
    \caption{
        Mean value of max splitting energy $\expval{E_s}$ (dot ``$\cdot$'') and its standard deviation $\sigma[E_s]$ (cross ``$\times$'') for (a) short wires; (b) long wires.
    }
    \label{fig:std_vs_mean}
\end{figure}

In Fig.~\ref{fig:std_vs_mean}, we show the mean value and standard deviation of maximal MZM splitting energy $E_s$ as a function of disorder strength $\sigma$ for short wires ($L=0.6,0.8,1~\mu$m) and long wires ($L=5,10,15~\mu$m). We find that as disorder strength $\sigma$ increases, the standard deviation of $E_s$ increases from 0 to a similar magnitude as the mean value of $E_s$, indicating a broad distribution of $E_s$ in the strong disorder regime.

\begin{figure*}[ht]
    \centering
    \includegraphics[width=6.8in]{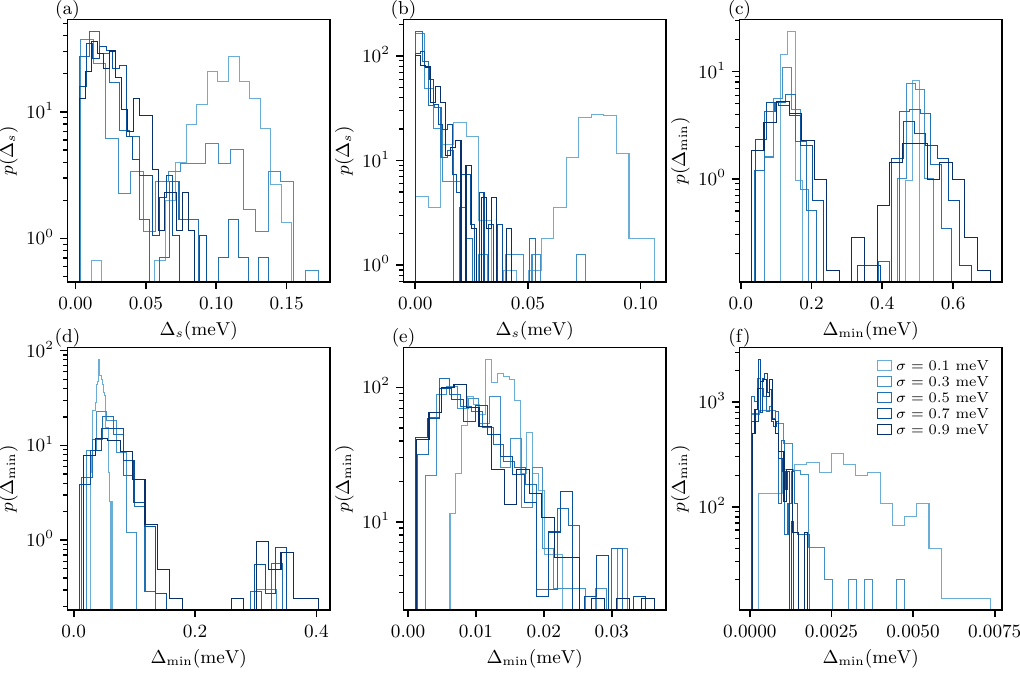}
    \caption{
        (a-b) The distribution of the gap size $\Delta_s$ at the same Zeeman field which maximizes the MZM splitting energy $E_s$ in the first lobe for (a) $L=3~\mu$m and (b) $L=10~\mu$m as a function of disorder strength $\sigma$.
        (c-f) The distribution of the minimal gap after gap reopens $\Delta_{\text{min}}$ for (c) $L=0.6~\mu$m, (d) $L=1~\mu$m, (e) $L=3~\mu$m, and (f) $L=10~\mu$m as a function of disorder strength $\sigma$.
    }
    \label{fig:gap_dist_ext}
\end{figure*}

In Fig.~\ref{fig:gap_dist_ext}(a-b), we show the distribution of the gap size $\Delta_s$ for $L=3$ and 10 $\mu$m in addition to the data for $L=0.6$ and 1 $\mu$m in the main text.
In Fig.~\ref{fig:gap_dist_ext}(c-f), we show distribution of the minimal gap after gap reopens $\Delta_{\text{min}}$, and gap size $\Delta_s$. 
Since the minimal gap can happen at any Zeeman field in the topological regime, making the splitting-over-gap ratio unbounded; therefore, we only focus on the same-Zeeman-field gap size $\Delta_s$ in the main text.

\begin{figure*}[ht]
    \centering
    \includegraphics[width=6.8in]{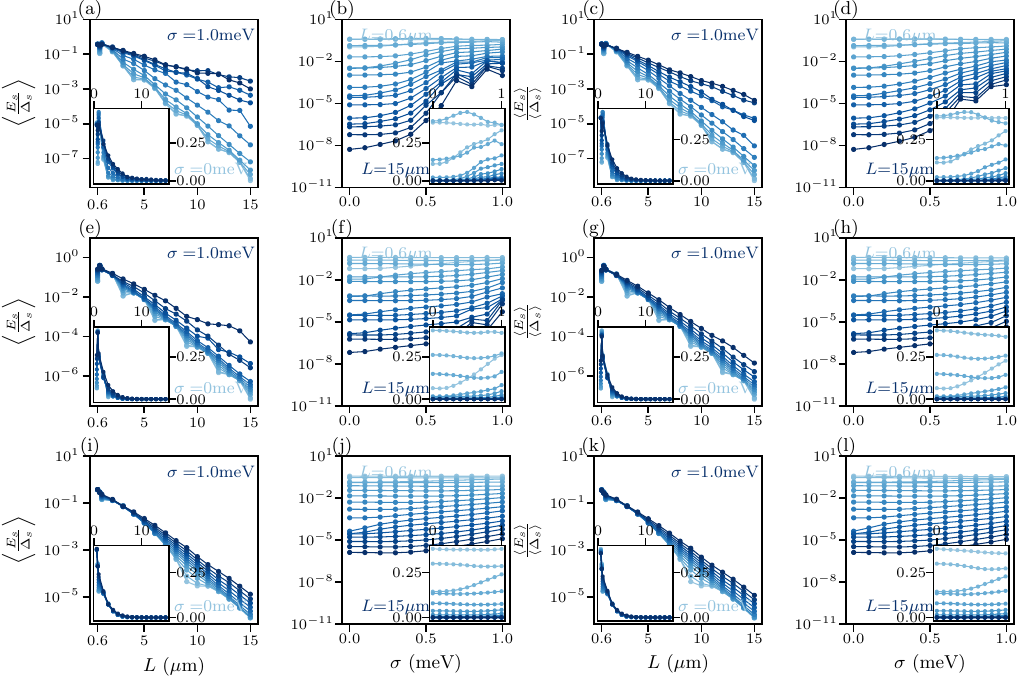}
    \caption{
        First row: Quenched average $R_1=\expval{\frac{E_s}{\Delta_s}}$ and annealed average $R_2=\frac{\expval{E_s}}{\expval{\Delta_s}}$ at $V_Z=1.2$ meV as a function $L$ or $\sigma$ in the semilog scale. 
        Second row: Same quantity but at $V_Z=1.5$ meV;
        Third row: Same quantity but at $V_Z=1.8$ meV. 
        Insets show the data in the raw scale.
    }
    \label{fig:ratio_ext}
\end{figure*}

In Fig.~\ref{fig:ratio_ext}, we show the quenched and annealed average of the ratio of maximal MZM splitting energy $E_s$ to gap size $\Delta_s$ at different Zeeman field $V_Z=1.2,1.5,1.8$ meV in addition to the data for the maximal MZM splitting energy in the main text.

\section{Superconducting coherence length}\label{app:coherence_length}
\begin{figure*}[ht]
        \centering
        \includegraphics[width=6.8in]{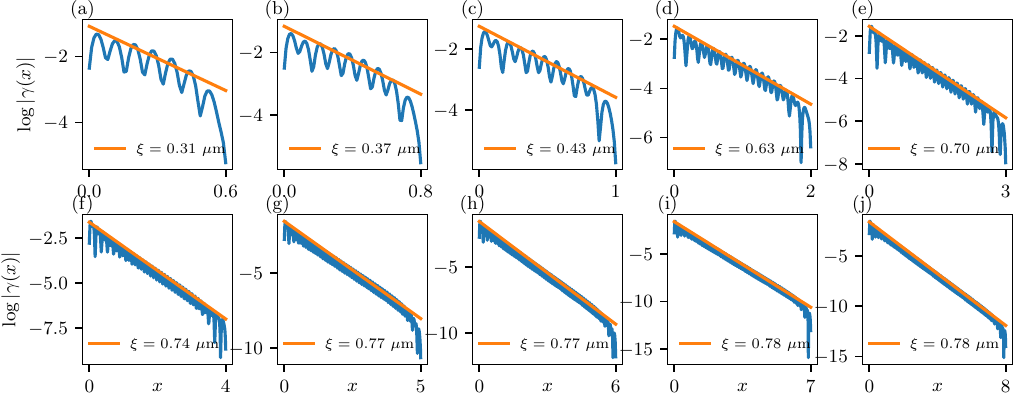}
        \caption{Wave function amplitude (blue curve) decays in the semilog scale in the pristine case, and its exponential fit (orange curve) for (a) $L=0.6~\mu$m, $V_z=2$ meV; (b) $L=0.8~\mu$m, $V_z=1.6$ meV; (c) $L=1~\mu$m, $V_z=1.3$ meV; (d) $L=2~\mu$m, $V_z=1.2$ meV; (e) $L=3~\mu$m, $V_z=1.1$ meV; (f) $L=4~\mu$m, $V_z=1.1$ meV; (g) $L=5~\mu$m, $V_z=1.2$ meV; (h) $L=6~\mu$m, $V_z=1.2$ meV; (i) $L=7~\mu$m, $V_z=1.2$ meV; (j) $L=8~\mu$m, $V_z=1.1$ meV. Other parameters are the same as in the main text.}
        \label{fig:example_pristine}
    \end{figure*}

\begin{figure}[ht]
        \centering
        \includegraphics[width=3.4in]{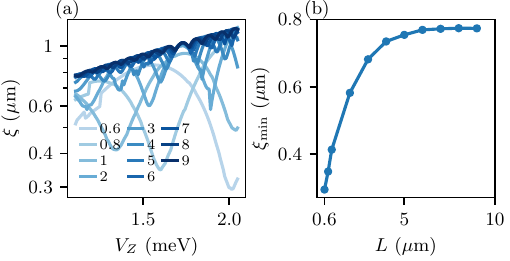}
        \caption{(a) The coherence length $\xi$ as a function of Zeeman field $V_Z$ for different wire length $L$ in the unit of $\mu$m; (b) The minimal coherence length $\xi_{\min}$ as a function of wire length $L$.}
        \label{fig:pristine_coh_len}
    \end{figure}

In this section, we extract the superconducting coherence length defined as 
\begin{equation}
    \abs{\gamma_{1,2}(x)} \sim e^{-\frac{x}{\xi}},
\end{equation}
where $\gamma_{1,2}(x)$ are the wavefunctions in the Majorana basis at the two ends of the wire.
We fit the wavefunction decay from the left end of the wire as shown in Fig.~\ref{fig:example_pristine} for different system sizes at a given $V_Z$.
The extracted coherence length $\xi$ as a function of $V_Z$ for different wire length $L$ is summarized in Fig.~\ref{fig:pristine_coh_len}(a).
For smaller $L$, the coherence length estimation is less accurate simply because the wire is not long enough to host well-separated MZMs. For very long wires, the coherence length will saturate to a constant, therefore it suffices to just go up to $L=10~\mu$m as shown in Fig.~\ref{fig:pristine_coh_len}(b).


\begin{figure*}[ht]
        \centering
        \includegraphics[width=6.8in]{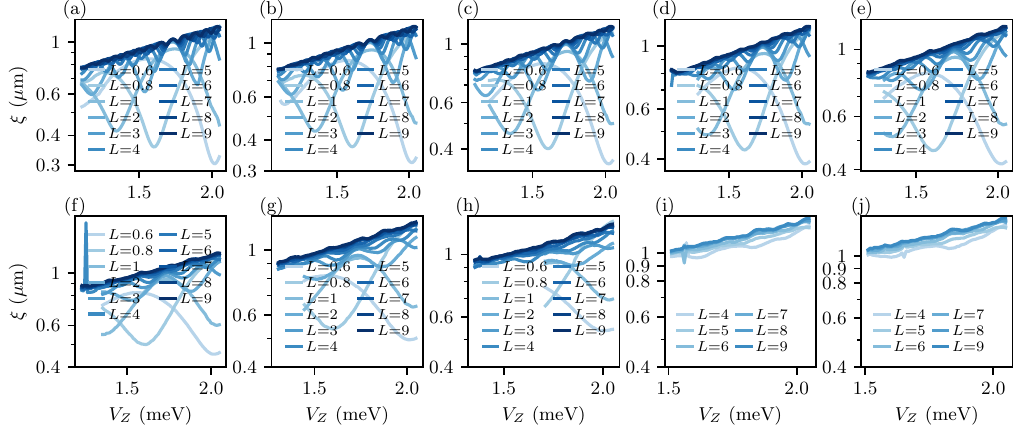}
        \caption{The coherence length $\xi$ as a function of Zeeman field $V_Z$ for different wire length $L$ and disorder strength $\sigma$. (a-j) $\sigma=0.1$ meV to 1 meV in steps of 0.1 meV.} 
        \label{fig:example_disorder}
\end{figure*}

\begin{figure}[ht]
        \centering
        \includegraphics[width=3.4in]{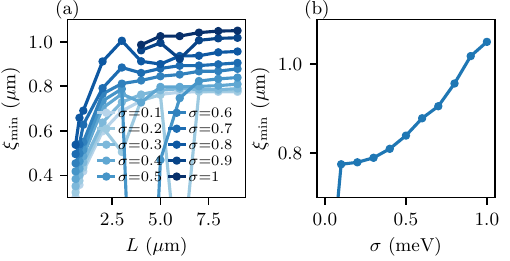}
        \caption{(a) The coherence length $\xi$ as a function of Zeeman field $V_Z$ for different wire length $L$ and disorder strength $\sigma$; (b) The minimal coherence length $\xi_{\min}$ as a function of disorder strength $\sigma$.}
        \label{fig:disorder_coh_len}
    \end{figure}

 With finite disorder, we extract the coherence length $\xi$ from the lowest-state wave function for each configuration of disorder realization and take an ensemble average, where the results are shown in Fig.~\ref{fig:example_disorder} for different system sizes and disorder strengths $\sigma$ as a function of $V_Z$. Here, we only show the data where the wave function amplitude decays exponentially. This is the reason why, for larger disorder strength, the extracted coherence length data start from $V_Z=1.5$ meV, which is much larger than the putative TQPT point $V_{Zc}=1$ meV in the pristine limit, and why it increases with increasing disorder strength, as shown in Fig.~\ref{fig:disorder_coh_len} (because they are effectively extracted at larger $V_Z$ for larger disorder strength). 

\section{Correlation between the splitting energy and topology}\label{app:correlation}

\begin{figure*}[ht]
        \centering
        \includegraphics[width=6.8in]{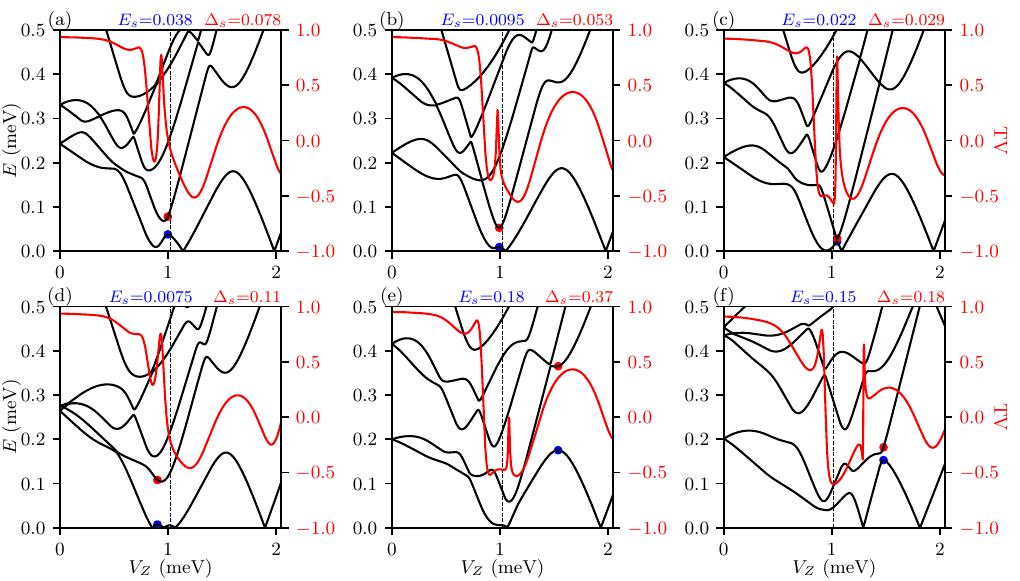}
        \caption{The band structure (black curves) and the topological visibility (red curves) as a function of Zeeman field $V_Z$ for (a-c) $\sigma=0.3$ meV; (d-f) $\sigma=0.9$ meV.
        } 
        \label{fig:correlation_gap_vs_TV}
\end{figure*}

In this section, we study the relationship between the splitting energy and the topological invariant. We follow the standard definition of the topological visibility (TV)~\cite{dassarma2016how} as the norm of the determinant of the transmission scattering matrix at zero energy.
We show the results in Fig.~\ref{fig:correlation_gap_vs_TV} for two different disorder strengths $\sigma=0.3$ meV and 0.9 meV. We find that the large gap size (shown in black curves) does not necessarily correspond to the most negative TV (shown in red curves).
More specifically, for small disorder these two quantities correlate more strongly, while for large disorder they do not correlate well.

\section{More examples of nanowire band structure}\label{app:bandstructure}
\begin{figure*}[ht]
    \centering
    \includegraphics[width=6.8in]{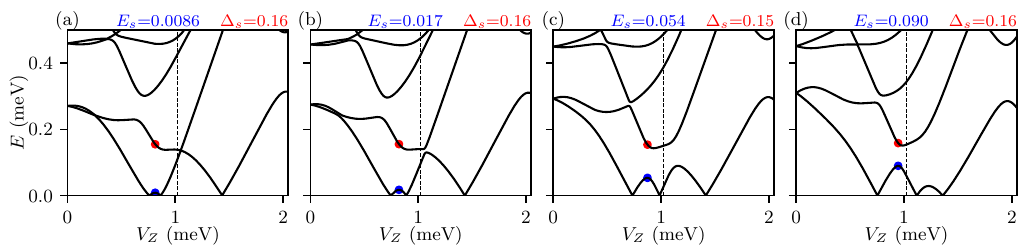}
    \caption{
        Band structure as a function of Zeeman field $V_Z$ for (a) pristine limit $\sigma=0$, (b) $\sigma=0.1$ meV, (c) $\sigma=0.5$ meV, and (d) $\sigma=0.9$ meV for a very short wire length $L=0.6~\mu$m.
        The blue (red) dot marks the maximal MZM splitting energy $E_s$ (gap size $\Delta_s$) with value indicated on the top of each panel.
    }
    \label{fig:bandstructure_L0.6}
\end{figure*}

\begin{figure*}[ht]
    \centering
    \includegraphics[width=6.8in]{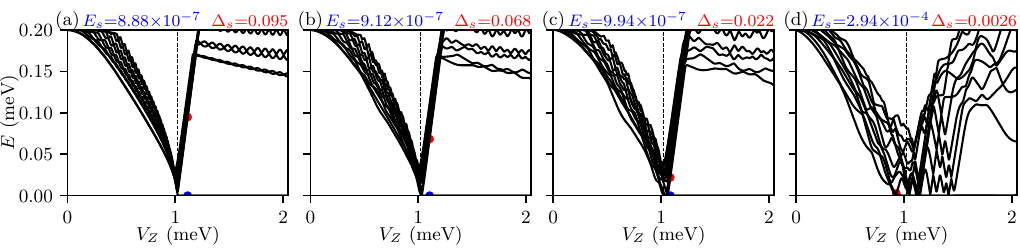}
    \caption{
        Band structure as a function of Zeeman field $V_Z$ for (a) pristine limit $\sigma=0$, (b) $\sigma=0.1$ meV, (c) $\sigma=0.5$ meV, and (d) $\sigma=0.9$ meV for a long wire length $L=10~\mu$m.
        The blue (red) dot marks the maximal MZM splitting energy $E_s$ (gap size $\Delta_s$) with value indicated on the top of each panel.
    }
    \label{fig:bandstructure_L10}
\end{figure*}

In this section, we present more examples of the band structure of the nanowire for a very short wire length $L=0.6~\mu$m in Fig.~\ref{fig:bandstructure_L0.6} and a long wire length $L=10~\mu$m in Fig.~\ref{fig:bandstructure_L10}.

\section{Nanowire model considering the self-energy of the superconductor}\label{app:self_energy}
In this section, we consider the self-energy correction in the proximity effect of proximitized-superconductivity in the nanowire~\cite{sau2010robustness,stanescu2010proximity}, which is a standard approach to model the SM-SC hybrid system, as discussed in many prior works~\cite{stanescu2014soft,stanescu2017proximityinduced,cole2015effects,liu2017andreev}.
We choose the same parameters as in the main text, except for the effective SC-SM coupling strength being 0.2 meV. Therefore, the TQPT remains at 1.02 meV.

\begin{figure*}[ht]
    \centering
    \includegraphics[width=6.8in]{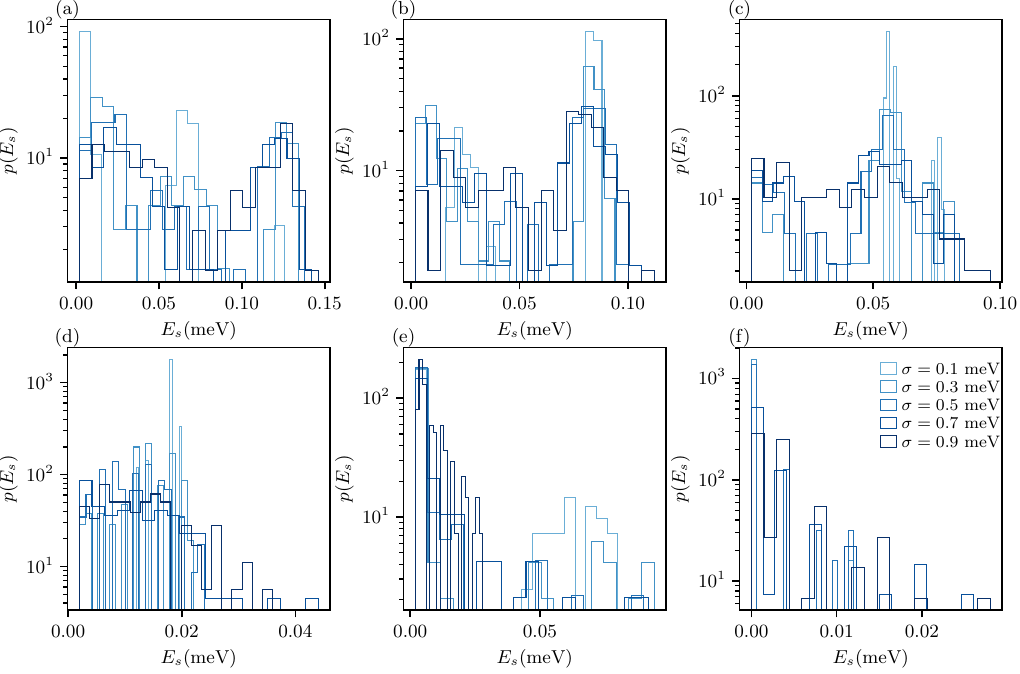}
    \caption{
        The distribution of maximal MZM splitting energy $E_s$ in the first lobe in a nanowire model with an effective SC-SM coupling of 0.2 meV as a function of disorder strength $\sigma$ for (a) $L=0.6~\mu$m, (b) $L=0.8~\mu$m, (c) $L=1~\mu$m, (d) $L=2~\mu$m,  (e) $L=3~\mu$m,  (f) $L=5~\mu$m.
    }
    \label{fig:Es_dist_SE}
\end{figure*}

\begin{figure}[ht]
    \centering
    \includegraphics[width=3.4in]{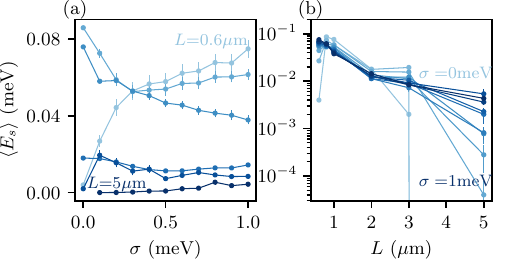}
    \caption{
        The distribution of the maximal MZM splitting energy $E_s$ in the first lobe in a nanowire model with an effective SC-SM coupling of 0.2 meV  (a) as a function of disorder strength $\sigma$ for different wire lengths $L$; (b) as a function of wire length $L$ for different disorder strengths $\sigma$.        
    }
    \label{fig:Es_scaling_SE}
\end{figure}

\begin{figure}[ht]
    \centering
    \includegraphics[width=3.4in]{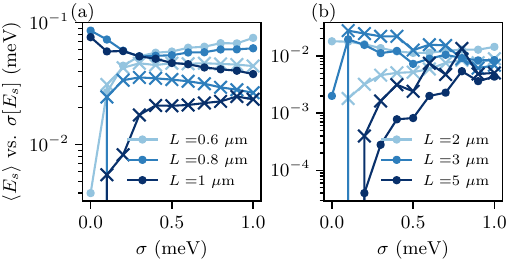}
    \caption{
        Mean value of max splitting energy (dot ``$\cdot$'') versus its median value (cross ``$\times$'') as a function of disorder strength $\sigma$ for short wires from $L=0.6$ to $3~\mu$m in a nanowire model with an effective SC-SM coupling of 0.2 meV.       
    }
    \label{fig:std_vs_mean_SE}
\end{figure}

We show the distribution of the maximal MZM splitting energy $E_s$ in the first lobe as a function of disorder strength $\sigma$ for different wire length $L$ in Fig.~\ref{fig:Es_dist_SE}.
We also show the scaling of the disorder-averaged maximal MZM splitting energy $\expval{E_s}$ as a function of disorder strength $\sigma$ and wire length $L$ in Fig.~\ref{fig:Es_scaling_SE}.
Finally, we show the mean value of maximal MZM splitting energy $E_s$ versus its median value as a function of disorder strength $\sigma$ for short wires from $L=0.6$ to $3~\mu$m in Fig.~\ref{fig:std_vs_mean_SE}.

\section{Kitaev chain}\label{app:kitaev_chain}
In this section, we study the Kitaev chain model~\cite{kitaev2001unpaired} in the presence of disorder, where the Hamiltonian is given by
\begin{equation}\label{eq:kitaev_chain}
    \begin{split}
        H &= -t\sum_{i=1}^{L-1} \left( c_i^\dagger c_{i+1} + h.c. \right) + \Delta \sum_{i=1}^{L-1}\left( c_i c_{i+1} + h.c. \right) \\
        & - \mu \sum_{i=1}^{L} c_i^\dagger c_i + \sum_{i=1}^{L} V_i c_i^\dagger c_i,
    \end{split}
\end{equation}
where $t$ is the hopping amplitude, $\Delta$ is the superconducting pairing amplitude, $\mu$ is the chemical potential, and $V_i$ is the disorder potential at site $i$.
We consider the disorder potential $V_i$ normally distributed with zero mean and standard deviation $\sigma$.
To compare with the SM-SC nanowire model in Eq.~\eqref{eq:H}, we set the parameters as $\Delta=0.2$, and $\mu=1$.
We vary the hopping amplitude $t$ and track the MZM splitting energy to locate its maximal splitting in the first lobe $E_s^{\text{K}}$.

\begin{figure}[ht]
    \centering
    \includegraphics[width=3.4in]{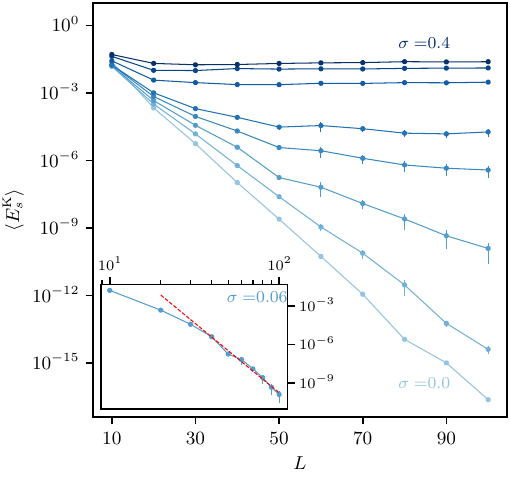}
    \caption{
        The disorder-averaged maximal MZM splitting energy $\expval{E_s^{\text{K}}}$ in the first lobe as a function of the wire length $L$ for different disorder strengths $\sigma$ (from light to dark blue: $\sigma=0,0.04,0.06,0.08,0.1,0.2,0.3,0.4$). 
        Lower-left inset: The power-law fit of $\expval{E_s^{\text{K}}}$ for $L\ge40$ at the crossover point $\sigma=0.06$ in the log-log plot showing an algebraic decay.}
    \label{fig:Es_vs_L_Kitaev}
\end{figure}

We notice that the averaged maximal MZM splitting energy $\expval{E_s^{\text{K}}}$ as a function of system size $L$ in the Kitaev chain as shown in Fig.~\ref{fig:Es_vs_L_Kitaev} also exhibits a crossover from exponential decay to power-law decay as the disorder strength $\sigma$ increases, where the crossover happens at a disorder strength $\sigma\approx 0.06$, which is smaller than the superconducting gap $\Delta=0.2$.

\begin{figure*}[ht]
    \centering
    \includegraphics[width=6.8in]{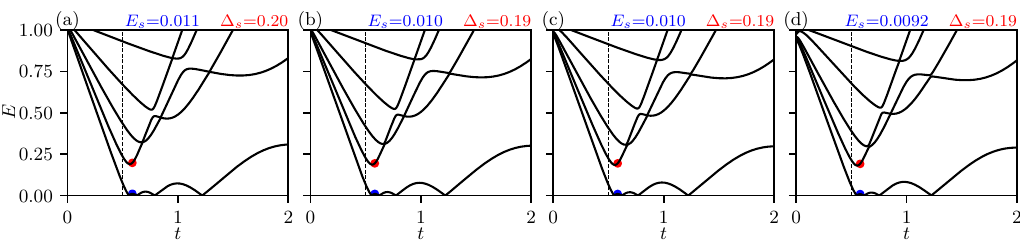}
    \caption{
        Band structure as a function of $t$ for (a) pristine limit $\sigma=0$, (b) $\sigma=0.04$, (c) $\sigma=0.06$, and (d) $\sigma=0.1$ for $L=10$.
        The blue (red) dot marks the maximal MZM splitting energy $E_s^{\text{K}}$ (gap size $\Delta_s^{\text{K}}$) with value indicated on the top of each panel.
    }
    \label{fig:bandstructure_Kitaev_L10}
\end{figure*}

\begin{figure*}[ht]
    \centering
    \includegraphics[width=6.8in]{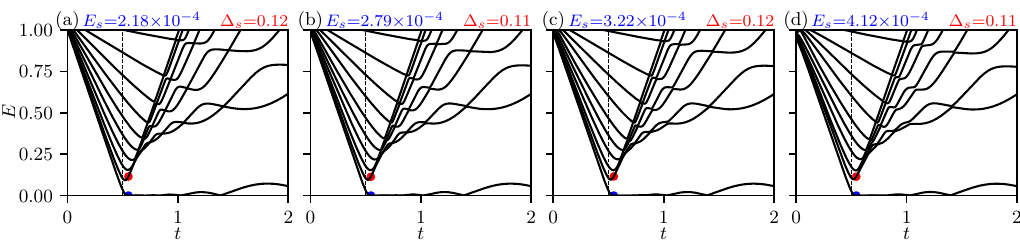}
    \caption{
        Band structure as a function of $t$ for (a) pristine limit $\sigma=0$, (b) $\sigma=0.04$, (c) $\sigma=0.06$, and (d) $\sigma=0.1$ for $L=20$.
        The blue (red) dot marks the maximal MZM splitting energy $E_s^{\text{K}}$ (gap size $\Delta_s^{\text{K}}$) with value indicated on the top of each panel.
    }
    \label{fig:bandstructure_Kitaev_L20}
\end{figure*}

\begin{figure*}[ht]
    \centering
    \includegraphics[width=6.8in]{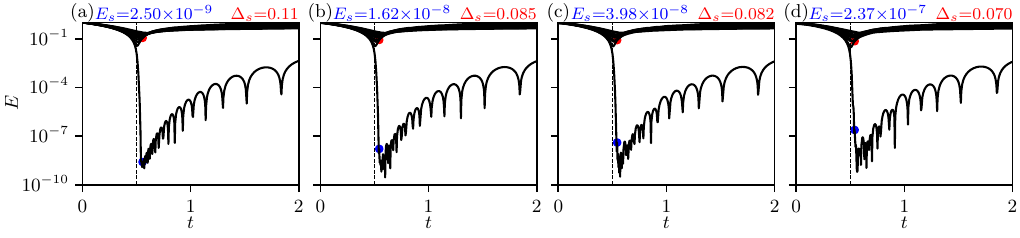}
    \caption{
        Band structure as a function of $t$ for (a) pristine limit $\sigma=0$, (b) $\sigma=0.04$, (c) $\sigma=0.06$, and (d) $\sigma=0.1$ for $L=50$.
        The blue (red) dot marks the maximal MZM splitting energy $E_s^{\text{K}}$ (gap size $\Delta_s^{\text{K}}$) with value indicated on the top of each panel.
    }
    \label{fig:bandstructure_Kitaev_L50}
\end{figure*}

We also present some typical examples of the band structure of the Kitaev chain as a function of $t$ for $L=$10, 20, and 50 in Fig.~\ref{fig:bandstructure_Kitaev_L10}, ~\ref{fig:bandstructure_Kitaev_L20}, and ~\ref{fig:bandstructure_Kitaev_L50}.

\end{document}